\documentstyle[psfig]{mn}
\title{A Parallel Tree-SPH Code for Galaxy Formation}
\author[Lia and Carraro]
       {
Cesario Lia$^{1}$ and Giovanni Carraro$^{2}$\\
       $^{1}$ SISSA/ISAS, Via Beirut 2, I-34014
	Trieste, Italy       \\
       $^{2}$ Dipartimento di Astronomia, Universit\`a di Padova, Vicolo
dell'Osservatorio 5, I-35122 Padova, Italy \\
E-mail: {\tt liac@sissa.it; carraro\char64pd.astro.it}
}

\date{\it Submitted: April 1999}
\pubyear{1999}
\begin{document}
\maketitle
\title{A parallel Tree-SPH code for Galaxy Formation}
%%%%%%%%%%%%%%%%%%%%%%%%%%%%%%%%%%%%%%%%%%%%%%%%%%%%%%%%%

%%%%%%%%%%%%%%%%%%%%%%%%%%%%%%%%%%%%%%%%%%%%%%%%%%%%%%%%%
\begin{abstract}
We describe a new implementation of a parallel Tree-SPH code with
the aim to simulate Galaxy Formation and Evolution.
The code has been parallelized using SHMEM, a Cray proprietary library
to handle communications between the 256 processors of the Silicon
Graphics
T3E massively parallel supercomputer hosted by the Cineca
Super-computing Center (Bologna, Italy).\\
The code combines the Smoothed Particle Hydrodynamics (SPH) method
to solve hydro-dynamical equations 
with the popular Barnes and Hut (1986) tree-code to perform gravity  
calculation with a $N \times logN$ scaling, and it is based on the
scalar Tree-SPH
code developed by Carraro et al (1998)[MNRAS 297, 1021].\\
Parallelization is achieved distributing particles along processors
according to a work-load criterion.\\
Benchmarks, in terms of load-balance and scalability, of the code are
analyzed and critically discussed against the
adiabatic collapse of an isothermal gas sphere test using $2 \times 10^{4}$
particles on 8 processors.
The code
results balanced at more than $95\%$ level. Increasing the number of
processors, the load balance slightly worsens.  
The deviation from perfect
scalability at increasing number of processors is almost negligible up to
32 processors.
Finally we present a simulation of the formation of an X-ray galaxy cluster
in a flat cold dark matter cosmology,
using $2 \times 10^{5}$ particles and 32 processors, and compare our
results with Evrard (1988) P3M-SPH simulations.\\
Additionally we have incorporated radiative cooling, star formation,
feed-back from  SN\ae\ of type II and Ia, stellar winds and UV flux from
massive stars,
and an algorithm to follow the
chemical enrichment of the
inter-stellar medium. Simulations with some of
these ingredients are also presented.
\par
\end{abstract}

\begin{keywords}
Methods: numerical --- Cosmology: theory --- Galaxy: formation 
\end{keywords}

\section{Introduction}

Galaxies and clusters of galaxies are believed to result from the gravitational instability 
of density fluctuations existing in the matter distribution of the primordial Universe.
These fluctuations  in the earliest phases grow linearly, and the evolution of the Universe 
in this linear or quasi-linear regime is generally studied by means of analytical tools
(Peebles 1993).
Galaxies and clusters however are clumpy and highly non-linear systems, and cannot be 
studied analytically; their formation and evolution must be modelled and followed numerically.
The standard procedure is to consider the matter density distribution emerging from different
cosmological models, and then to simulate the non-linear regime of structure formations
using numerical simulations.
The most widely accepted cosmological scenario is the hierarchical Cold Dark Matter (CDM)
model, which is based on two ideas. The first one is that the Universe for most of its life has 
been  dominated by  an unknown kind of collision-less material, called Dark Matter (DM), and 
the second one is  that the structures growth proceeds hierarchically
(White \& Rees 1978), the less massive objects
forming before the most massive ones, which are assembled through the merging of smaller more 
ancient objects.
 The so-called Dark era has been widely studied starting from the pioneering work of Davis et
al (1985), who compared the DM spatial distribution emerging from N-body simulations, 
with the galaxies catalogues available at  that time. 
When more powerful computers were available, the properties of individual galaxy halos
have been studied in great detail, concluding that independently from the cosmological model,
the primordial density fluctuations spectrum and the halo mass, all the halos show uniform
universal properties in their matter distribution (Navarro et al 1996,
Huss et al 1998, Moore et al 1998).

However on galaxies scale the evolution is governed not only by the DM, but also by the gas,
whose dynamics regulate on large scale the formation  of grand design spiral arms and extended 
thin disks, and on smaller scale the formation of stars and the interaction between stars and the 
multi-phase interstellar medium (ISM) (Thornton et al 1998).
To understand how galaxies formed and evolved it is therefore necessary to couple 
gravitational forces and hydro-dynamics .
This can be done semi-analitycally or numerically.
Semi-analytical models of galaxy formation (Kauffmann et al 1993), 
although successful in reproducing many galaxy 
properties (Baugh et al 1998), are based on several ad-hoc assumptions about the interaction 
between gas and dark matter, and the way stars form from gas and interact with the ISM.
Many physical processes like thermal shocks, pressure forces and dissipation are required,
together with gravity,  to realistically model the formation and evolution of galaxies.
This can be done more properly by means of numerical simulations, in which the equations of motion 
of a large numbers of particles and/or grid cells are integrated under 
gravitational and hydro-dynamical forces.

In numerical astrophysics gravity can be computed using different methods,
which can generally be divided in particle and grid technique.  
The simplest particle method ($PP$) directly sums up the pairwise contribution between all the particles,
but has the disadvantage to increase the computational time as $N\times N$, making impossible to handle
simulations with more than $10^{4}$ particles (Aarseth 1985).  Although this method  can be used
for other purposes (for example, open star clusters, Aarseth 1998), it cannot be appropriate for testing cosmic
structure and/or galaxy formation  theories.  A recent development is represented by GRAPE (GRAvity
PipE) boards, where the $PP$ force computation is performed with a special-purpose hardware (Hut \& Makino 1999).
Tree codes (Barnes \& Hut 1986,  Appel 1985)  reduce the scaling to $N\times logN$ 
putting the system in a hierarchical structure, and computing the forces from distant particle 
groups in an approximate fashion.
The Particle-Mesh ($PM$) method is based on a grid evaluation of the newtonian gravitation potential  
using Fast Fourier Transforms (FFT). It scales as $N\times logN$ as well, but has the drawback that 
the spatial
resolution is limited by the cell size (Hockney \& Eastwood 1981).
The resolution limitation can be alleviated using a hybrid method, called $P^{3}M$, which uses the grid
estimate of the potential only for distant regions, whereas the neighboring contributions are
computed with the $PP$ method. However the best way to circumvent resolution problems is to
use adaptive grids, which deform and adapt to regions of different density  (Hydra, Pearce 
and Couchman  1997).

Hydro-dynamical forces are calculated adopting a Lagrangian or Eulerian formalism.
The use of grids is natural for Eulerian codes. In  these codes the values of  hydro-dynamical 
quantities are estimated  inside the grid cells, whereas fluxes are evaluated across the cell
borders, as in the Piecewise Parabolic Method (PPM) (Woodward \& Colella 1984; Brian
et al 1995, Gheller et al 1998a).
The spatial resolution can be  improved using Adaptive Mesh Refinement (AMR, 
Bryan \& Norman 1995).
On the other hand, Lagrangian codes start from the hydro-dynamical conservation laws in
Lagrangian formalism (Landau \& Lifchitz 1971), and mostly utilize particles to map the fluid
properties. This is the case of  the Smoothed Particle Hydro-dynamics (SPH) technique
developed by Lucy (1977) and Gingold \& Monaghan (1977). The advantage of this technique
is the great flexibility and adaptivity (Hernquist \& Katz 1989; Steinmetz \& M\"uller 1993;
Carraro et al 1998a).
 
In the last decade a variety of different combinations of gravity solvers and 
hydro-dynamical methods appeared. Briefly,  there are pure grid codes like $PM-PPM$
(Bryan \& Norman 1995), grid codes combined with particle codes like $P^{3}M-SPH$ 
(Evrard 1988), or pure particle codes, like GRAPE-SPH (Steinmetz 1996) or Tree-SPH 
(Hernquist \& Katz 1989). Basically Lagrangian particle codes provide a better resolution
in the over-dense regions, but exhibit a poorer shocks resolution and under-dense regions
resolution compared with Eulerian codes (Kang et al 1994).
However most  astro-physical phenomena occur in high density regions, in particular
galaxies and clusters are over-dense regions. This fact renders lagrangian code more favourite
to study astro-physical problems, provided the involved dynamics is not 
strongly dominated by shocks.

Carraro et al (1998a) developed a pure particle code, combining Barnes \& Hut (1986) octo-tree
with SPH,  and applying this code to the formation of a spiral galaxy like the Milky Way.
The code is similar to Hernquist \& Katz (1989) TreeSPH. It uses SPH to solve the 
hydro-dynamical equations
(see also Carraro et al 1998b;
Lia \& Carraro 1999). In SPH a fluid is sampled using particles, there is no resolution
limitation due to the absence of grids, and great flexibility thanks to the use of a time and space
dependent smoothing length.  
Shocks are captured by adopting an artificial viscosity 
tensor,  and
the neighbors search is performed using the octo-tree. The octo-tree, combined with SPH, 
allows a global time scaling  of $N\times logN$.  A good advantage of such codes is that it is easy to
introduce new physics, like cooling and radiative processes, magnetic fields and so forth.
Finally the kernel, which is utilized to perform hydro-dynamical 
quantities estimates,  can be made adaptive by using anisotropic smoothing lengths 
(Shapiro et al 1996). 

It is widely recognized that TreeSPH codes, although deficient
in some aspects,  can give reasonable answers in many astrophysical situations, like in simulations
of fragmentation and star formation in giant molecular clouds (GMC)
(Monaghan \& Lattanzio 1991), supenov\ae~ explosions 
(Bravo \& Garcia-Senz 1995),  merging of galaxies (Mihos \& Hernquist 1994), galaxies
and clusters formation (Katz \& Gunn 1991, Katz \& White 1993) 
and  Lyman alpha forest (Hernquist et al 1996).
 
Galaxy formation in particular requires a huge dynamical range (Dav\'e et al 1997). In fact
an ideal galaxy formation simulation would start from a volume as large as the universe to
follow
the initial growth of the cosmic structures, and  at the same time would be able to resolve
regions as small as GMC, where stars form and drive the galaxy evolution through their
interaction with ISM. This ideal simulation would encompass a dynamic range of $10^{9}$
(from Gpc to parsec), $10^{6}$ time greater than that achievable with present day codes.

Big efforts have been made in the last years to enlarge as much as possible the dynamical range 
of  numerical simulations, mainly using more and more powerful supercomputers:
scalar and vector computers indeed cannot handle efficiently a number of particles greater than
half a million (Katz et al 1996).\\
A successful example
is the Virgo Consortium (Glanz 1998), which  has been 
recently able to perform  simulations of the Hubble Volume (a cube of
$2000h^{-1} Mpc$ on a side
on  the Cray T3E
supercomputer by using  a number of particles  close to $10^{9}$.  They used a 
parallelized $P^{3}M-SPH$ code.

Dav\'e  et al (1997) for the first time developed a parallel implementation of a TreeSPH
code (PTreeSPH) which can follow both collision-less and collisional matter.
They report  results of simulations run on a Cray T3D computer of the adiabatic collapse 
of an initially isothermal gas sphere (using 4096 particles),  of  the collapse of a 
Zel'dovich pancake (32768 particles) and of a cosmological simulation 
(32768 gas and 32768 dark particles).

Their results are quit encouraging, being quite similar to those obtained with the scalar 
TreeSPH code (Hernquist \& Katz 1989). 
Porting a scalar code to a parallel machine is far from being an easy task.  A massively 
parallel computer  (like the Silicon Graphics T3E) links together hundreds 
or thousands of processors 
aiming at increasing  significantly the computational power. For this reason they are very 
attractive, although several  difficulties can  arise in adapting a code to these machines.
Any processor possesses its own memory, and can assess other processors memory by
means of communications which are handled by a hard-ware network, and are usually slower than 
the computational speed. 
Great attention must be paid to avoid situations in which a small number of processors
are actually working while most of them are standing idle. Usually one has to invent
a proper data distribution scheme which allows to subdivide particles into processors
in a way that any processor handles about the same number of particles and does not
need to make heavy communications. Moreover the computational load must be
shared between processors, ensuring that  processors exchange informations
all together, in a synchronous way, or that  any  processor is performing different 
kinds of work  when it is waiting for information coming from other processors, in an 
asynchronous fashion (Dav\'e et al 1997).

In this paper we present a parallel implementation of the TreeSPH code described in Carraro
et al (1998a). The numerical ingredients are the same as in the scalar version of the code.
However the design of the parallel implementations required several changes to the
original code.
The key idea that guided us in building the parallel code was to avoid continuous
communications, limiting the information exchange at a precise moment along the code flow.
This clearly reduces the communication overhead. 
We have also decided to tailor the code to the machine, improving its efficiency. 
Since we are using a T3E massively parallel computer, a natural choice was to handle 
communications using the SHMEN libraries,  which permit asynchronous communications,
and are intrinsically very fast, being produced  directly by Cray for the T3E super-computer (Lia \& Carraro 1999).
At present the code is also portable to other machine, like SGI Origin 2000, and may be
portable to any other machine thanks to the advent of the second release of  Message Passing Interface
(MPI).  

The plan of this work is as follows. Section 2 is devoted to a brief description of the 
scalar code, whereas in Section 3 we highlight the parallelization
strategy. In section 4 and 5 we describe the adiabatic collapse of an isothermal
gas sphere, while in Section 6 we discuss
the code performances in terms of load balance and scalability. 
Section 7 shows simulations
including cooling of gas, 
whereas in section 8 we present a cosmological
Nbody/hydro simulation of the formation of a galaxy cluster.
Finally section 9 summarizes our results and discusses future work.

\section{The TreeSPH code}
The parallel Tree-SPH code we are going to  present is derived from the scalar Fortran 90
Tree-SPH code described in Carraro et al (1998a). It is a combination of the SPH  
method (see Monaghan 1992 for an exhaustive review) and the Barnes \& Hut (1986)
tree-code.  

\subsection{The tree-code}
The octo-tree developed by Barnes \& Hut (1986) encompasses all the system under study in
a cubic box, called root. Then  the root is subdivided into 8 cells, each one with its own sample 
of particles. This subdivision proceeds down cell by cell iteratively until each sub-cell contains 
1 or no particles at all. The building of this tree structure can be done putting one particle at a time
(bottom-up) or sorting the particles across divisions (top-down). Both methods scale as $N \times logN$.  
For any cell the total mass, center of mass and higher order multipole moments (usually up to the
quadrupole order) are computed. Typically the tree is built up at any time-step , since it is a very
rapid operation.

Once the tree is built up, it is possible to calculate the force on a particle "walking" down
the tree level by level starting from the root. The basic idea, which renders this scheme very
fast, is to approximate the forces from distant particles. This is done using an opening criterion.
Briefly,  at each level an interaction list is made, which contains cells if they are sufficiently distant,
or particles, if the cells are close. In this case in fact the cells are opened, looking at their content.
A widely used version of the opening criterion is

\begin{equation}
d > \frac{l}{\theta} + \delta
\end{equation}

\noindent
which is derived from Barnes (1994).
In this equation $l$ is the cell size, $d$ is the distance of  a particle from the cell center of mass,
$\delta$ is the distance from the cell center of mass and the cell geometric center and, finally,
$\theta$ is the opening angle. This criterion, which replaces the classical one,

\begin{equation}
d > \frac{l}{\theta} 
\end{equation}

\noindent
guarantees to avoid pathological situations when the center of mass is close to the cell border
(Dubinski 1997).

To obtain  the force on a particle it is necessary to loop through the accumulated interaction
list , and this loop represents the real amount of computation. It is evident that the interactions 
number is much smaller than in the classical PP method. Dubinski (1997) calculates that on 
average there are about 1000 interactions per particle in a simulation with one million particles.

The value of $\theta$ is somewhat
arbitrary, only at decreasing $\theta$ values the number of  openings increases and the forces are
more accurate. Hernquist (1987) showed that using  $\theta  = 1$ implies errors on the particles
accelerations amounting to $1\%$.

Gravitational interactions are then softened to avoid numerical divergences when two particles
get very close. This is done introducing a softening parameter $\epsilon$ which corresponds to
attribute a dimension to the particles. We decide to use a Plummer softening, equal for all
the particles, computed looking at the inter-particles separation in the central regions
of the system under investigation (Romeo, 1997).

In the TreeSPH code developed by Carraro et al (1998a) the Barnes \& Hut (1986) treecode has
been included in the code as a subroutine.

\subsection{SPH overview}
SPH is a method to solve the hydro-dynamical conservation laws in Lagrangian form, 
which has been shown by Monaghan (1992) to be an example of an interpolating technique.
The fluid under study is sampled using particles, and the hydro-dynamical quantities are
estimated at particle positions. This is done smoothing each particle physical properties
with a kernel ( Gaussian,
exponential or spline) over some smoothing lengths, and deriving gas properties adding 
up the contribution from a number  of neighbors.

Benz (1990) in a famous review showed how to derive the SPH form of the hydro-dynamical 
equations. They read:

\begin{equation}
\rho_i = \sum_{j}  m_j W(\vec r_i - \vec r_j, h_i,h_j)
\end{equation}

\begin{equation}
\frac{d \vec v_i}{dt} = -\sum_{j}m_j(2\frac{\sqrt(P_i P_j)}{\rho_i \rho_j } + \Pi_{ij}) \nabla_i W
\end{equation}

\begin{equation}
\frac{d u_i}{dt } = \sum{_j}m_j (\frac{\sqrt(P_i P_j)}{\rho_i \rho_j } + \frac{1}{2}
\Pi_{ij}) (\vec v_i - \vec v_j) \nabla_i W .
\end{equation}

\noindent
In these equations $h$ is the particle smoothing length, which is estimated according to the particle
density (Benz 1990),  and $m$, $v$, $P$, $u$ and $\rho$ are the mass, velocity, pressure, specific internal energy
and density associated with each particle.
The kernel W has been taken from Monaghan \& Lattanzio (1985), and it is a spline kernel with 
compact support within 2 smoothing lengths:

\begin{equation}
W(r,h) = \frac{1}{\pi h^{3}} \left\{ \begin{array}{ll}
1 - \frac{3}{2} u^{2} + \frac{3}{4} u^{3}, & \mbox{if $0 \leq u
\leq 1$;}\\
\frac{1}{4} (2 - u)^{3},                   & \mbox{if $1 \leq u
\leq 2$;}\\
0.                                         & \mbox{otherwise,}
                                     \end{array}
                             \right.
\label{kspl}
\end{equation}

The kernel is then symmetrically averaged according to Hernquist \& Katz (1989):

\begin{equation}
W_{ij} = \frac{1}{2} \cdot (W_i + W_j) 
\end{equation}

\noindent
This guarantees momentum conservations.
Neighbors are searched for walking down the tree, and looking at those gas particles which actually 
are within  2 smoothing lengths.\\
The tensor $\Pi_{ij}$ is the viscosity tensor,  introduced to
treat thermal shocks:

\begin{equation}
\Pi_{ij} = \left\{ \begin{array}{ll}
\frac{-\alpha c_{ij} \mu_{ij} + \beta \mu_{ij}^{2}}{\rho_{ij}}, &
\mbox{if $(\vec v_{i} - \vec v_{j}) \cdot (\vec r_{i} - \vec r_{j})
>
0$} \\
0.                                                              &
\mbox{otherwise,}

                 \end{array}
           \right.
\end{equation}
\noindent
where

\begin{equation}
\mu_{ij}=\frac{h_{ij}({\vec v}_i-{\vec v}_j)({\vec r}_i-{\vec
r}_j)}{|{\vec
r}_i-{\vec r}_j|^2 + \epsilon h_{ij}^2}  .
\end{equation}

\noindent
Here $c_{ij}=0.5(c_i+c_j)$ is the sound speed,
$h_{ij}=0.5(h_i+h_j)$, and $\alpha$ and $\beta$
are the viscosity parameters, usually set to 0.5 and 1.0,
respectively.
The factor
$\epsilon$ is fixed to 0.01 and is meant to avoid divergencies.

As amply discussed by Navarro \& Steinmetz (1997) this formulation 
has the disadvantage of not vanishing in the case of shear dominated
flows, when $\vec \nabla \cdot \vec v = 0$ and $\vec \nabla \times
\vec v \neq 0$. In such a  case,   a spurious shear viscosity can
develop, mainly in simulations involving a small number of particles.
To reduce the shear component we adopt the Balsara (1995)
formulation of the viscosity tensor

\begin{equation}
\tilde \Pi = \Pi_{ij} \times \frac{f_{i} + f_{j}}{2},
\end{equation}

\noindent
where $f_i$ is a suitable function defined as

\begin{equation}
f_i=\frac{|<{\vec \nabla} \cdot  {\vec v}>_i|}{|<{\vec \nabla} \cdot {\vec
v}>_i| + |<{\vec \nabla }\times {\vec v}>_i|+\eta c_i/h_i},
\end{equation}  

\noindent
and $\eta \approx 10^{-4}$ is a parameter meant to prevent
numerical divergencies.

Time steps are calculated according to the Courant condition 

\begin{equation}   
\Delta t_{C,i} = {\cal C} \frac{h_{i}}{h_{i} \left | \vec \nabla \cdot \vec
v_{i}
\right | + c_{i} + 1.2(\alpha c_{i} + \beta max_{j} \left | \mu_{ij}
\right |)}   ,
\end{equation}

\noindent
with $\cal C$ $\approx 0.3$.

In presence of  gravity, a more stringent condition on  the time steps is
required. According to Katz, Weinberg \& Hernquist (1996), the additional
criterium has to be satisfied

\begin{equation}
\Delta t_{G,i} = \eta \cdot MIN (\frac {\eta \epsilon}{|\vec v |},
(\frac{\epsilon}{|\vec a|})^{1/2}),
\end{equation}

\noindent
where $\epsilon$ is the gravitational softening parameter and
 $\eta$ is another parameter usually set to 0.5.
The final time step
to be adopted is the smallest of the two

\begin{equation}
\Delta t_{i} = MIN (\Delta t_{C,i},\Delta t_{G,i}).
\end{equation}

We adopt the multiple time-steps scheme. This is done building up
a binary time-bins structure, as discussed by Hernquist \& Katz (1989).
Hydro-dynamical forces and gravity are then computed only for active
particles (i.e. those particles which are occupying the smallest time bin), 
while for the remaining particles they are interpolated
(position, velocity, internal energy and potential)
or extrapolated (density and smoothing length).
With active particles we mean those particles which occupy
the lowest time-bin, and for this reason evolve more fastly.

\subsection{A standard Tree-SPH time-step}
Having described the basics of our Tree-SPH code,
it is necessary to show what the code precisely does in a single
$n-th$ time-step.  
The leap-frog integration scheme, for the positions and the velocities, 
proceeds as follows:

Firstly an estimate of the velocity $\tilde {\vec v}_i^{n+1/2}$ is obtained
from

\begin{equation}
\tilde {\vec v}_i^{n+1/2}= {\vec v}_i^n +0.5 \Delta t_{i} {\vec a}_i^{n-1/2}.
\end{equation}

\noindent
This is used to compute time-centered accelerations ${\vec 
a}_i^{n+1/2}$,
walking down the tree and performing SPH estimates of the hydro-dynamical forces.
Then  particle velocities and positions  are
updated. 
 
\begin{equation}
{\vec v}_i^{n+1}={\vec v}_i^n +\Delta t_{i} {\vec a}_i^{n+1/2}
\end{equation}

\begin{equation}
{\vec r}_i^{n+1/2}={\vec r}_i^{n-1/2}+\Delta t_{i} {\vec v}_i^n   .
\end{equation}

\noindent
The energy equation is explicitly solved, unless sink or source terms
are present, and particle energies are advanced in the same manner
as positions. We will return to
the specific internal energy integration in Section~7.1.

\section{The parallel TreeSPH code}

Our parallel code has been implemented for the massively parallel computer
Silicon Graphics T3E hosted by the CINECA Super-computing
Center (Bologna, Italy).
It consists of 256 DEC Alpha 21164 processors, 128 cpus having
128 Mbyte and 128 cpus having  256 Mbyte of RAM memory, the total memory
amounting to about 49
Gbyte.
Peak performances are evaluate at 300 GygaFLOPS.
The processors are inter-connected with
a hardware  network, which allows to transmit 480 Mbyte/sec.
In developing the parallelization strategy, we opted to keep the code
tightly tailored to the machine, with the aim to increase as much as
possible its performances. 

\subsection{Thinking in Parallel}
The advent of massively parallel computers offers the opportunity
to significantly reduce the time one has to wait for the results of
a simulation involving just one parameter set, provided it is
straightforward to port a code in a parallel machine.
This is not necessarily true.\\
A parallel computer in fact is a collection of individual workstations,
each one with its own cpu and memory (at least the NUMA - Non Uniform Memory
Access - class), 
which can access other workstations
information, or data, only through communications, which obviously
take some time.
Therefore to obtain a real speed-up it is
crucial that the data the code is processing are well distributed
within the processors in order to evenly share the amount of computation
between processors. If this situation is maintained during all
the simulation (no processor stands idle for much time waiting for 
receiving data),
the decrease of the time necessary to perform a simulation
can be enormous, also taking into account the time spent
exchanging information.\\
The first step in porting a scalar code into a parallel computer
is to identify which part of the computation is local, and
which remote. With local computation we mean the computation which
does not require information from the other processors, because all the
necessary data are available locally. With remote we mean those parts
of a code that cannot proceed without receiving data from distant
processors.\\
Using this terminology, the calculation of the gravitational potential
is heavily remote, because to advance a particle, it is necessary
to sum up the contribution to the force acting on it from all the other particles.
On the other hand, the SPH calculation is remote as well, although
in this case communications are ideally restricted to the surrounding
processors, those ones which actually possess particles inside 2 smoothing
lengths.
Completely local are for instance the computation of cooling and heating
processes and  star formation, which usually are done particle by
particle.\\
Having recognized the local and remote parts of the code, there
are two options:

\begin{description}
\item [$\bullet$] Any cpu can retrieve each time it needs remote data from
a remote cpu;
\item [$\bullet$] Any cpu can grab all the data it needs from the other cpus
in a single burst of communication.
\end{description}

\noindent
The choice between these two possibilities must be carefully weighted
at the porting stage.\\
The first solution is obviously the simplest one, since the coding is
straight-forward and moreover there are semi-automatic tools (like HPF-CRAFT)
which manage the communications during the compilation phase. In some cases
(Gheller et al 1998b) this guarantees good load-balance and acceptable efficiency.
However the easy implementation hiddens the real role of communications 
(Sigurdsson et al 1997), which unfortunately are crucial in a Tree-SPH code.
The lack of any communication control would lead to a strong
code degradation, since the amount of computation is mostly remote,
the code would spend most of the time communicating and the speed-up would
be almost  negligible.

A suitable data and work distribution must be implemented, and this is
the topic of the next two sections. \\
To handle communication we decided to use the SHMEM libraries, which
are advantageous in a number of ways. They have been released by the
former Cray to make the best use of the T3E hard-ware communication system.
Therefore these libraries intrinsically ensure fast communications.
Moreover they allow the use of an asynchronous communication scheme,
that is to say two cpus which want to exchange information, do not have
to be necessarily synchronized. This for instance permits to a cpu to perform
some computation in the meantime it waits for some data to arrive,
provided these data are not necessary to proceed further.

\subsection{Data Structure}
As discussed above a crucial step in parallel coding
is to work out a suitable data distribution
scheme. 
Dubinski(1997) and Dav\`e et al (1997) algorithm stands on the Salmon (1991) Recursive 
Orthogonal Bisection Rectangular Domains Decomposition scheme. Any cpu handles
a portion of the overall volume, whose size is weighted on the amount of local work.
Briefly, the idea is to keep track of the particles computational work 
(work-load), and to distribute the particles in domains with roughly the same amount
of computational load.
This does not necessarily mean that the sizes of the domains are equal,
neither that the number of particles inside any cpu is the same.\\
The work load can be estimated in several ways,
and the guiding criterion must be decided looking at those parts of the code
which actually tend to suffer more easily from unbalancing problems.
In a Tree-SPH code the number of interactions is a function of the particles density,
so an obvious solution would be to estimate the work-load as the inverse of the time-step.
For instance we obtained acceptable results using the following recipe:

\begin{equation}
load_{i} = const \cdot \frac{1}{t_{i}^{2}}
\end{equation}

\noindent
where $const$ is a user-defined parameter, and $t_i$ is the global particle time-step.
However better results can be obtained replacing the total time-step 
with the time spent to compute the gravitational force acting on a particle, 
in a manner similar
to Dav\`e et al (1997).\\ 
Going into some details, we firstly compute for any cpu the bary-center of the particles
work-load. 
Then the entire system work-load barycenter is estimated.
Axis by axis 
the simulation volume is recursively cut along a direction passing through the global
barycenter and orthogonal to the axis.

\begin{figure}
\centerline{\psfig{file=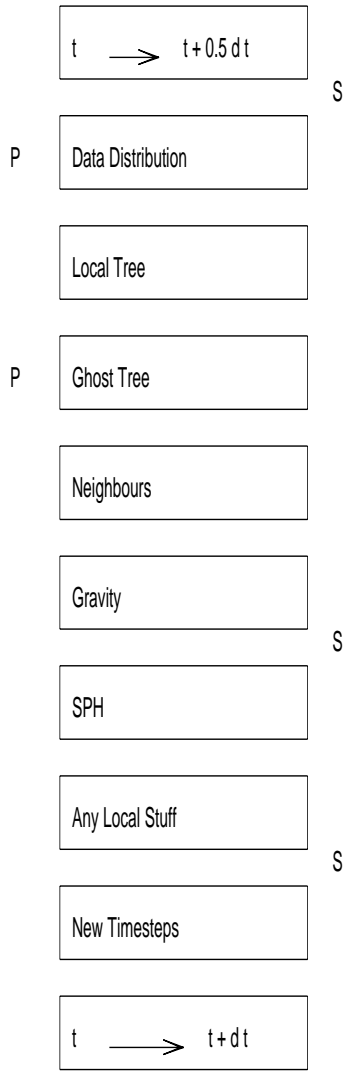,height=16cm,width=9cm}}
\caption{Schematic view of a single typical time step in a TreeSPH
code. Label P indicates the parallel parts of the code, whilst label S indicates
the synchronization points.}
\end{figure}

\subsection{The Parallel Tree-code}
The computation of the gravitational potential, as discussed above,
proceeds through three steps:

\begin{description}
\item[$\bullet$] building up the tree data structure;
\item[$\bullet$] definition of mass and center of mass of any cell;
\item[$\bullet$] walking down the tree level by level according to the opening
criterion to compute gravitational acceleration.
\end{description}

\noindent
Let us assume for a moment that all the data have been distributed
in some way between the processors (see Section 3.2).
The first two steps are accomplished by any cpu independently, without
any information exchange. In other words, any cpu builds up its local
tree with 
the local particles, and performs the calculation of mass and
center of mass up to the quadrupole terms.\\
Before computing forces (third step), any cpu walks the tree of all
the other cpus level by level according to the opening criterion. If
the criterion is satisfied for a remote cell, this cell is copied and
added to the local tree in what we call hereinafter the local
{\it ghost-tree}. This is clearly the analogous of the {\it Local Essential Tree}
of Dubinski (1997).\\ 
We must stress that the inspection of the remote trees is actually done
only if the opening criterion is satisfied for at least one active local particle,
being gravity calculated exactly only for these particles (see the discussion
in Section 2.2).
The building of the {\it ghost-tree} requires a unique burst of data
exchange, and actually represents a parallel stage in the code. Once the
{\it ghost-tree} is constructed, any cpu has all the necessary data
to compute particles acceleration walking down the local
{\it ghost-tree}, and proceeds without any further
communication.

\subsection{Parallel SPH}
In SPH the hydro-dynamical quantities at the location of a particle
are estimated - as discussed in Section 2.2 -
summing the contributions from  a sample of neighbors
smoothed with a kernel. This means that every active particle must keep
a list of its neighbors, which can reside in the same cpu, or in other
remote cpus. The basic idea again is to gather in a single communication
operation the neighbors necessary to evaluate density and specific
internal energy.
Dav\'e et al (1997) compile a list of locally essential particles (LEPL)
checking whether non local particles close to the cpu borders are within
two smoothing lengths.\\
In our case we decided to use the {\it ghost-tree} to perform neighbors
searching looking for particles within two smoothing lengths.
In such a way, in a unique burst of communications any cpu
collects all the necessary information to compute gravity and
hydro-dynamical forces.\\
A possible source of inconsistency is related to the necessity to
have at any time-step the synchronized values of the hydro-dynamical
quantities in order to make predictions.
While this is automatically guaranteed in a scalar code, it is not necessarily
ensured in a parallel one. In fact in our code all the data exchange occurs
during the building of the {\it ghost-tree}. This implies that
all the collected data are synchronized at the previous time-step. 
To better describe the situation, let us 
imagine that one cpu is going to SPH estimate the new specific internal
energy of  its particles. 
It must loop over the neighbors, which could reside either in the local tree
or in the {\it ghost tree}.
To up-date the energy, we need 
the up-dated values of density, pressure and sound speed for all the
neighbors. If all the neighbors are inside the local tree
the up-dated energy is automatically correct, otherwise it keeps the previous
not synchronized value.\\
In PTree-SPH (Dav\`e et al 1997) the problem is solved 
by means of a data exchange whenever
critical hydro-quantities have been locally up-dated.
This is done keeping  memory of the
cpu from which any neighbor comes, by using some kind of tagging.\\
The drawback of this solution is that
at least two supplementary communication operations are necessary.
One after the local smoothing length calculation, and the other soon
after the local density up-dating.
Although simple, this solution would deteriorate the code by increasing 
the over-head also in the case of an asynchronous communications scheme, 
like the one adopted by us. 
In fact we should synchronize two times all the cpus
by resorting to a BARRIER command.\\
We found that it is possible to circumvent this over-head problem at expense of 
some accuracy loss. 
This is done by extrapolating locally densities and smoothing
lengths (using the velocity divergence, Benz 1990 - see discussion in Section~2.2)  
not only for non active particles, but for all the particles. This 
would ensure synchronism. 
The extrapolated estimates would be obviously
different from the remote up-date values. 
Nevertheless these differences
do not affect significantly the results of our code, as the next sections
will show.

\begin{figure*}
\centerline{\psfig{file=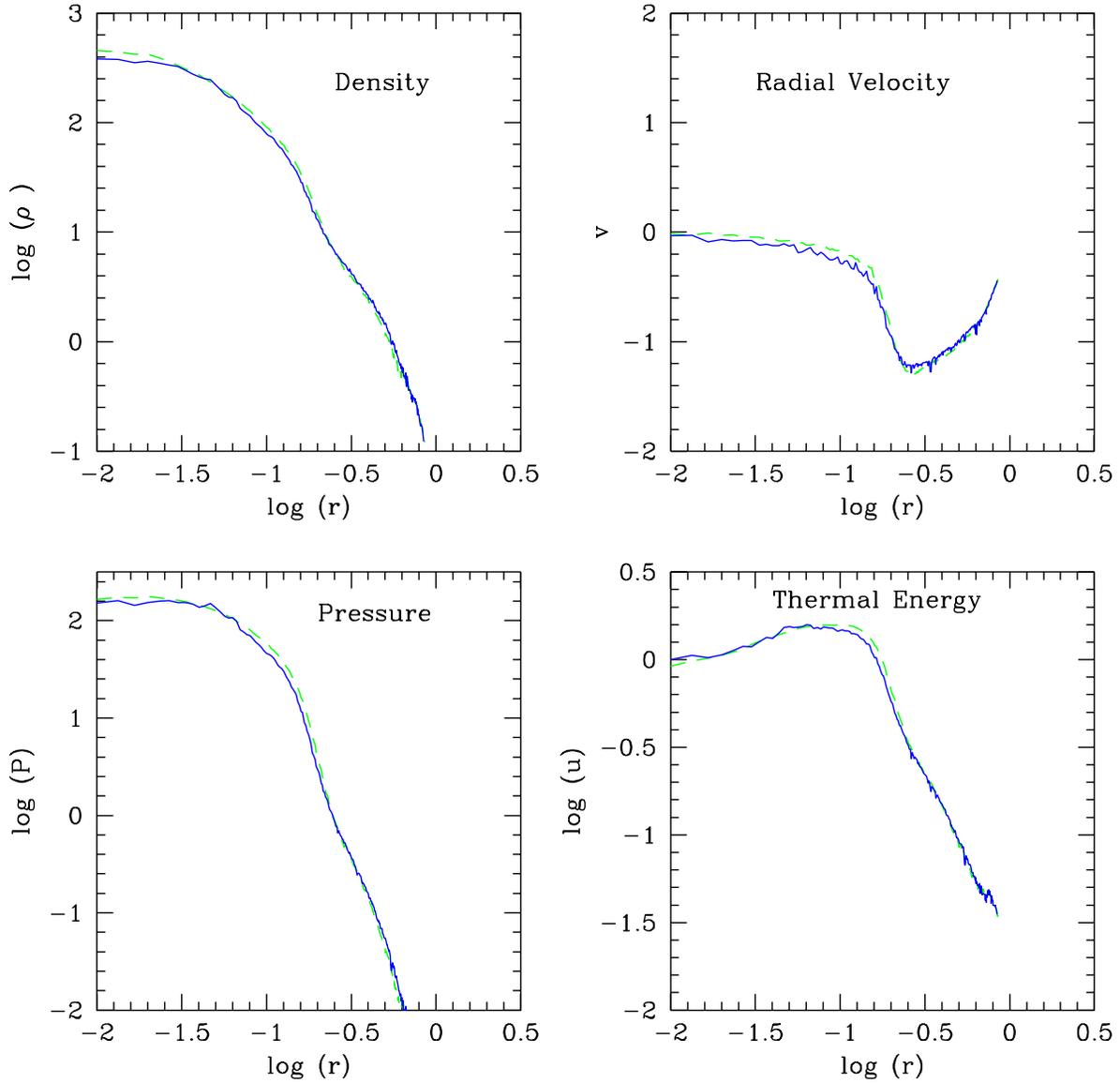,height=16cm,width=16cm}}
\caption{Adiabatic collapse: snapshots of the density, radial velocity,
pressure and internal energy at the time of the maximum compression. 
The results of the test  
performed using $2\times 10^{4}$ particles with 8 processors are shown
with dashed lines. Solid lines show the results obtained with the same
number of particles, but using the scalar 
code, for comparison.}
\end{figure*}

\subsection{Parallel over-head}
The use of a parallel machine necessarily produces some
over-head due to the data exchange and 
the processors synchronization.\\
In our code data communication is significantly reduced thanks
to the use of the {\it ghost-tree}, which allows to exchange data in a
single burst,  and of the SHMEM libraries, as discussed in Section~3.1.\\

The real source of over-head are the synchronizations (see Fig.~1).
Generally speaking, a single Tree-SPH step requires three
synchronization operations: during the {\it ghost-tree} building,
after the evaluation of density, pressure and sound speed, and
at the definition of the new time-step hierarchy.\\
The first  synchronization is less than a problem, since at the beginning
of the computation the processors are still well synchronized.
We circumvent the second one extrapolating the hydro-dynamical quantities
instead of performing a new data exchange (see Section~3.4).
The third synchronization is really a problem, since it is at the end of
the computation, where all the  unbalance goes  accumulating.
To better analyze this point, it is worth recalling that most of the
computing time is spent calculating the gravitational and hydro-dynamical
forces, and that it is done only for the subsample of particles
(the so called active particles),
which are occupying the smallest time bins, and are actually synchronized
with the entire system (see Section~3.2).\\
It seems reasonable to conclude that a good work balance could
be  obtained defining a work-load criterion only looking at the active
particles.
However this can often lead to severe memory problems, for instance any
time  the number
of active particles is a small fraction of the total number of particles.
In this case in fact most of the particles might be  handled by a few
processors.\\
To avoid such problems we decided to use in any case the criterion 
discussed in Section~3.2.

\section{Testing the code}
In this section we are going to present some applications of our code,
aiming at testing its capability to reproduce known results, and to
estimate its efficiency. To start, we present the adiabatic collapse
of an initially isothermal gas sphere,
a standard test for hydro-dynamical codes.\\
In the following sections we shall  discuss the collapse
of a mix of gas and DM, with and without cooling. Being cooling calculation
local - as discussed above -, this will allow us
to verify how the code reacts to the inclusion
of new physics, which does not affect the overall communications, but only
the amount of local computation.\\
Finally we study the formation of an X-ray cluster of galaxies
in a flat ($\Omega=1.0$) cold dark matter universe, from $z=4$ to
the present, with the aim of comparing our code with a similar test
performed with a P3M-SPH code by Evrard (Evrard 1988,1990).

\section{The adiabatic collapse of an isothermal gas sphere}

\subsection{Introduction}
We consider the adiabatic collapse of an 
initially non-rotating
isothermal gas sphere. This is a standard test for  SPH codes
(Hernquist \& Katz 1989;   
Steinmetz \& M\"uller 1993; Nelson \& Papaloizou 1994). 
In particular it is an ideal test for a parallel code,
due to the large dynamical range and high density contrast.
To facilitate the  comparison of  our results with those by the
above authors, we adopt the same
initial model and the same units ($M=R=G=1$).
The system consists of a $\gamma = 5/3$ initially isothermal gas sphere, with a 
density profile:

\begin{equation}
\rho(r) = \frac{M(R)}{2\pi R^{2}} \frac{1}{r}  ,
\end{equation}

\noindent
where M(R) is the total mass inside the sphere of radius R.
Following Evrard (1988), the density profile is obtained stretching an 
initially regular cubic grid by means of the
radial transformation 

\begin{equation}
r_{i}^{old} \Rightarrow r_{i}^{new} =
(\frac{r_{i}^{old}}{R})^{3}R.
\end{equation}

Alternatively it is possible to use the 
acceptance-rejection procedure as in  Hernquist \& Katz (1989).
The  total number of particles used in this simulation
is $2\times 10^{4}$. All the particles have the same mass.
The specific internal energy is set to $u = 0.05GM/R$.
For this test the viscosity parameters $\alpha$ and $\beta$
adopted are 1 and 2, respectively, in agreement with Dav\'e et a (1998).
The gravitational softening parameter $\epsilon$ adopted for
this simulation is $5 \times 10^{-3}$.

\begin{figure}
\centerline{\psfig{file=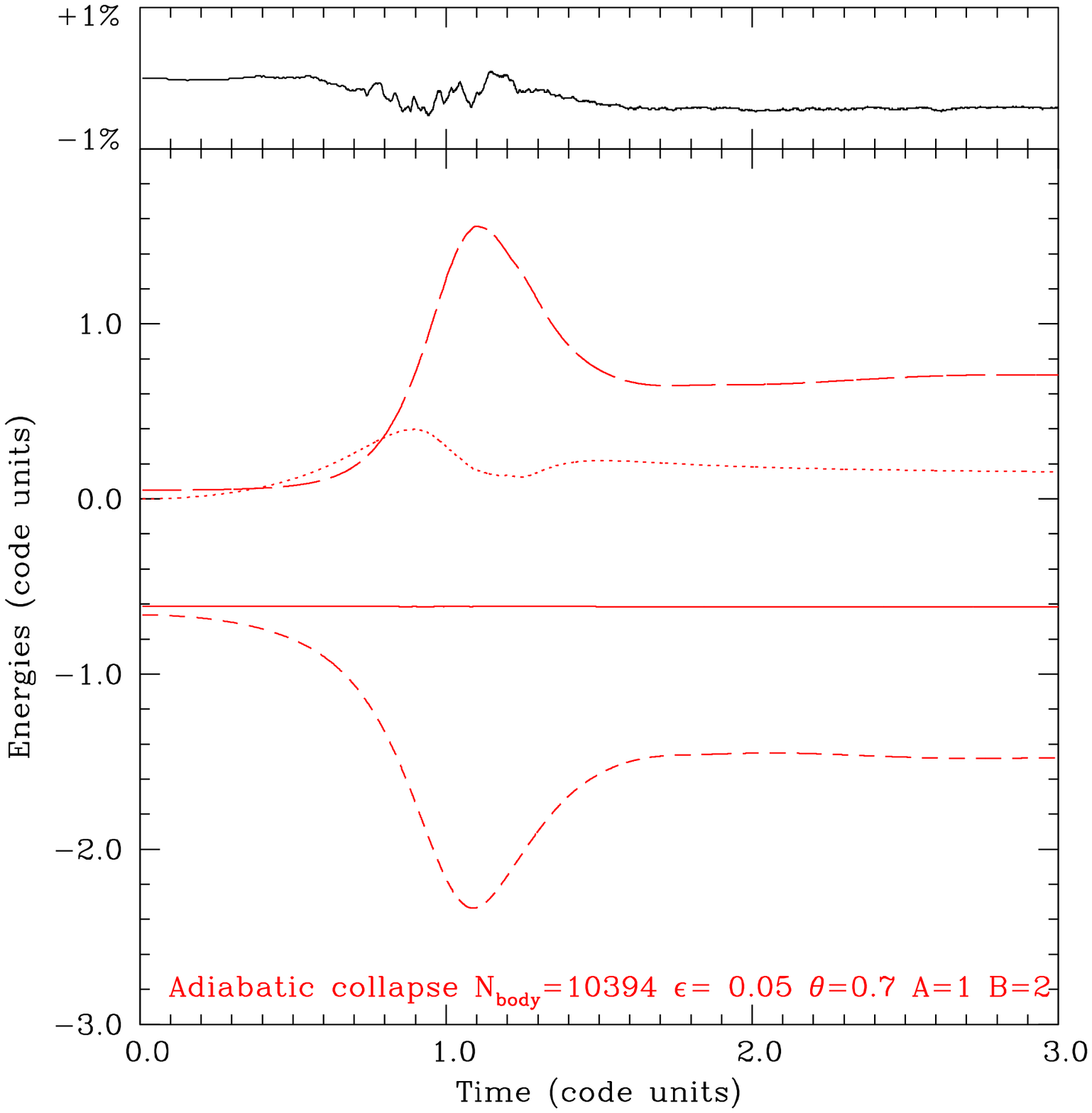,height=9cm,width=9cm}}
\caption{Energy trend in the adiabatic collapse.
The dashed-dotted line indicates potential energy, 
the dotted line kinetic, the dashed line thermal, and 
the solid line total energy. The top panel is a zoom
on the trend of total energy, showing the deviation
from perfect conservation.}
\end{figure}

%%%%%%%%%%%%%%%Table 1
\begin{table*}
\tabcolsep 0.4truecm
\caption{The Adiabatic Collapse test.
Benchmarks for a run with $2\times 10^{4}$ particles. Time refers
to 50 time-stpes.}
\begin{tabular}{cccccccc} \hline
\multicolumn{1}{c}{$N_{cpu}$} &
\multicolumn{1}{c}{Total} &
\multicolumn{1}{c}{Data Up-date} &
\multicolumn{1}{c}{Parallel Over-head} &
\multicolumn{1}{c}{Neighbors} &
\multicolumn{1}{c}{SPH} &
\multicolumn{1}{c}{Gravity} &
\multicolumn{1}{c}{Miscellaneous} \\
\hline
& secs &secs & secs & secs & secs & secs & secs\\
\hline
 1  &120    &0.47  &0.00    &40    &36    &40   &3.53\\ 
 2  &69     &0.22  &0.60    &23    &19    &25   &1.18\\ 
 4  &42     &0.27  &1.70    &14    &9.5   &15   &1.53\\ 
 8  &23     &0.13  &3.20    &5.5   &5.4   &5.3  &3.47\\ 
16  &17.3   &0.13  &3.40    &3.4   &3.0   &3.8  &3.60\\ 
32  &11.5   &0.09  &3.00    &2.6   &1.3   &3.2  &1.31\\
64  &7.5    &0.05  &2.90    &0.33  &1.1   &1.9  &1.23\\ 
\hline
\hline
\end{tabular}
\end{table*}

\subsection{Description of the tests}

The state of the system at the time of the maximum compression
is shown in the various panels  of Fig. 1, 
which displays the density, radial velocity, pressure  and specific internal 
energy profiles. Each panel shows the variation of the physical
quantity under consideration (in suitable units) as a function of the
normalized radial coordinate at time equal to 0.88 . 
  
The initial low internal energy is not sufficient to support the gas 
cloud which starts to collapse. Approximately after 
one dynamical time scale a bounce 
occurs. The system afterwards can be described as an isothermal core plus an 
adiabatically expanding envelope 
pushed by the shock wave generated at the stage of 
maximum compression. 
After about three dynamical times the system reaches virial equilibrium
with total energy equal to a half of the gravitational potential energy
(Hernquist \& Katz 1989).
The temporal evolution of the kinetic, thermal and potential energies,
is shown in Fig.~2. The trends are quite similar to Steinmetz \& M\"ller
(1993) and Hernquist \& Katz (1989).
Total energy conservation, measured
as the maximum deviation from
the perfect conservation, is ensured
below $1\%$.

The present results agree
fairly well with the mean values of the Hernquist \& Katz (1989) 
simulations, which in turns agree with the 1D finite difference
results (Evrard 1988). As an
example, the shock is located at the radial distance  $0.18 \leq r/R
\leq 0.25 $ in the models at $t \approx 0.88$ (cf. the velocity panels
of Fig.1) .
The good agreement between the results of the parallel and scalar tests
guarantees that the processors exchange data correctly.
The level of precision can be gauged by estimating the effect of the interpolation 
stage (which we use to avoid an additional communication flow) on the evaluation of the
thermal energy. To this aim we run the adiabatic collapse test with the scalar and parallel
(8 cpus) code, by imposing that all the particles are active. This indeed represents 
the most non uniform  situation. It turns out that the deviation due to the interpolation
amounts at maximun to $0.8\%$. The maximum deviation is located at the time of the maximum
compression (t~ $\approx$ ~0.88).

\begin{figure*}
\centerline{\psfig{file=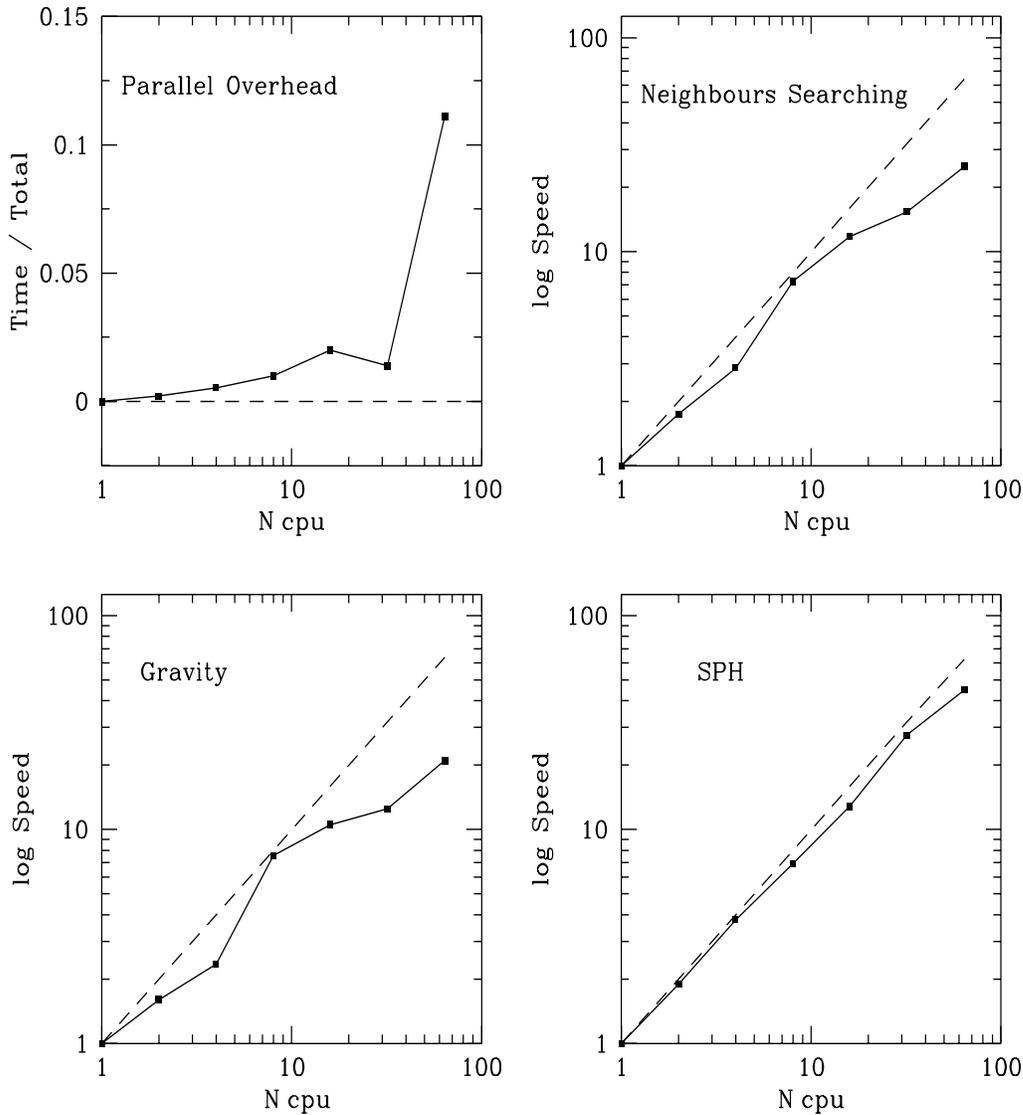,height=16cm,width=14cm}}
\caption{Scalability of different code sections (solid lines), as reported
in each panel, averaged
on 50 time-steps. Dashed lines indicate the ideal scalability.}
\end{figure*}

\section{Benchmark}
To evaluate the code performances, we use the adiabatic collapse just
described, and perform simulations at increasing number of processors.
We believe that this test is very stringent, and can give a lower 
limit of the code
performances due to the high density contrast that is present at the time
of maximum compression, when the particles are highly clustered. 
We are going to check the code timing, overall load-balance and scalability.
Moreover we shall analyze in details particular sections of the code,
like the gravity computations,the SPH and the neighbor searching.
An estimate of the parallel over-head will be given as well.

\subsection{Timing analysis}
We run the adiabatic collapse test up to the time of the maximum compression
(t $\simeq 1.1$) using $2 \times 10^{4}$ particles
on 1, 2, 4, 8, 16, 32 and 64 processors, and looked at the performances in
the following code sections (see also Table~1): 

\begin{description}
\item[$\bullet$] total wall-clock time;
\item[$\bullet$] data up-dating;
\item[$\bullet$] parallel computation,
which consists of barriers, the construction of the {\it ghost-tree} and the
distribution of data between processors;
\item[$\bullet$] search for neighbor particles;
\item[$\bullet$] evaluation of the hydro-dynamical quantities;
\item[$\bullet$] evaluation of the gravitational forces;
\item[$\bullet$] miscellaneous, which encompasses I/O and statistics.
\end{description}

\noindent
The results summarized in Table~1 present the total  
wall-clock time per processor over the last 50 time-steps,
together with the time spent in each of the 5 subroutines
(data updating, neighbor searching, SPH computation, gravitational
interaction and parallel computation). 
The gravitation interaction takes about one-third of the total time,
while the search for neighbors takes roughly comparable time.
The evaluation of hydrodynamical quantities (see Section 3.5) takes about
one-fourth of the time, the remaining time being divided between I/O and
data up-dating. The parallel over-head  does not appear to be a serious problem, being
at maximum  about $1\%$ of the total time.
This timing refers, as indicated above, to simulations stopped at roughly the time
of maximum compression. A run with 8 processors up to $t \simeq 2.5$, the time at which
the system is almost completely virialized, 
took 3800 secs. 
Global code performances are analyzed in the next sections.

\begin{figure}
\centerline{\psfig{file=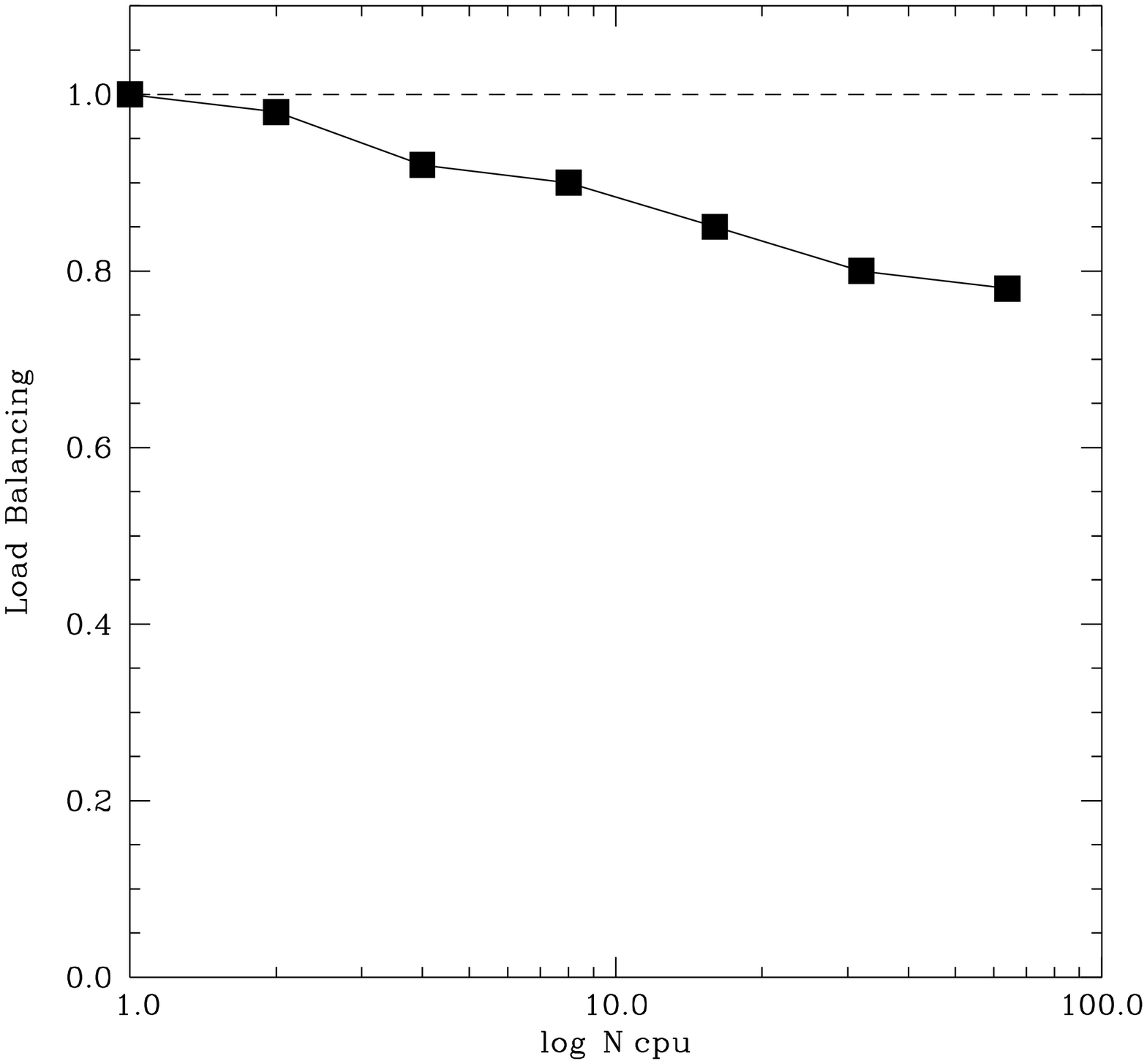,height=9cm,width=9cm}}
\caption{Overall code load-balance, averaged on 50 time-steps (solid line).
Dashed line indicates ideal load-balance.}
\end{figure}

\begin{figure}
\centerline{\psfig{file=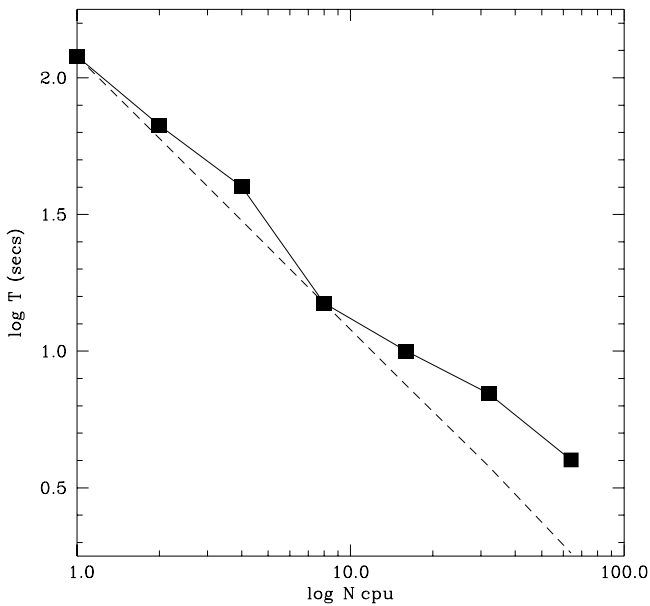,height=9cm,width=9cm}}
\caption{Overall code scalability, averaged on 50 time-steps (solid line).
Dashed line indicates ideal scalability.}
\end{figure}

\begin{figure}
\centerline{\psfig{file=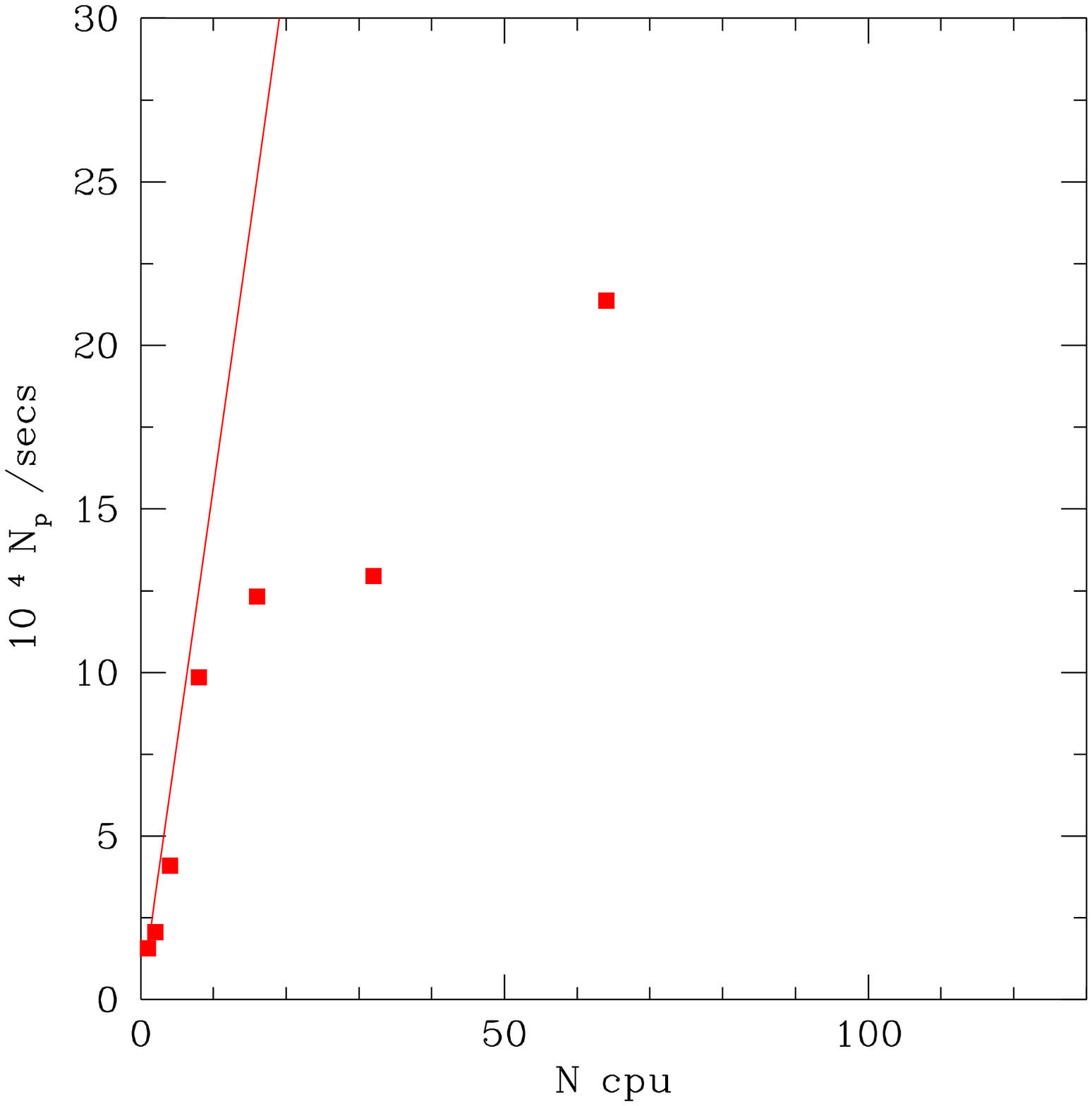,height=9cm,width=9cm}}
\caption{Code scalability in a pure collisionless simulation.
See text for more details.}
\end{figure}

\subsection{Load-balance}
One of the most stringent requirements for a parallel code is the capability to distribute
the computational work equally between all processors.
This can be done defining a suitable work-load criterion, as discussed in Section~3.2.
This is far from being an easy task (Dav\`e et al 1997), and in practice some
processors stand idly for some time waiting that the processors with the heaviest
computational load accomplish their work. This is true also when an asynchronous communication
scheme is adopted, as in our TreeSPH code.
As outlined in Section~3.2, we are using individual work-loads, based on the time spent 
to evaluate the gravitational interaction on one particle with all the other ones. 
A better choice would be to define the work-load only for active particles, which are 
the particles evolving fatly. This possibility is currently under investigation, due to
the memory problems that can arise, as discussed in Section~3.5.
To evaluate the code load-balance we adopted the same strategy of Dav\`e et al (1997),
measuring the fractional amount of time spent idle in a time-step 
while another processor performs computation:

\begin{equation}
L = \frac{1}{N_{procs}}\sum_{j=1}^{N_{procs}} 1 - \frac{(t_{max - t_i})}{t_{max}}   .
\end{equation}

Here $t_{max}$ is the time spent by the slowest processor, while $t_i$ is
the time taken by the $i-th$ processors to perform computation.
The results are shown in Fig~5, where we plot the load-balance for simulations
at increasing number of processors, from 1 to 64.
The load balance maintains always above $80\%$, being close to 1 up to 8 processors.
For the kind of simulations we are performing, the use of 8 processors is particularly
advantageous for symmetry reasons. 

\subsection{Scalability}
At increasing number of processors, a parallel code should ideally speed up linearly 
In practice the increase of the processors number causes an increase of the communications
between processors, and a degradation of the code performances.
To test this, we used the same simulations discussed above, running the adiabatic collapse 
test with $2 \times 10^{4}$ particles at increasing processors number. 
We estimated how the code speed scales computing
the wall-clock time per processor 
spent to execute a single time-step, averaged over 50 time-steps. In Fig.~6 we plot the 
speed (in $sec^{-1}$) against the number of processors. \\
The code scalability keeps very close to the ideal scalability up to 8 processors, 
where it shows a minimum. This case in fact is the  most symmetric one
Then the scalability deviates significantly only 
when using more that 16 processors. Looking also at Fig.~4, it is easy to recognize that 
mainly the gravitational interaction is responsible for this deviation.\\

To better judge the code performances, we run a simulation of the collapse
of a pure DM system, aiming at showing the scalability of the gravity section of the 
code. The results are shown in Fig.~ 7.  They are good up to 16 processors, afterwards
they suddenly get worse. This trend does not change by introducing all the other
code parts, as it will be shown in the next sections. This is clearly imputable 
to the dominant role of the gravity, which represents not only the most time consuming
section of any TreeSPH code (this holds also for the serial code), but also
by definition the less parallel part of the code.\\

In the next Sections we are going to investigate whether the code overall performances
might improve adding new physics
(like cooling and star formation) which is necessary to describe the evolution of real systems,
like galaxies.

\subsection{Memory considerations}
Together with the efficiency, considerations on the memory use of a parallel code are 
crucial (Dav\`e et al 1998).
A good use of the memory can significantly reduce the communication overhead. In fact, 
by increasing the number of particles per processor, the local computation increases with 
respect to the remote one, decreasing the amount of time spent commnicating. In our case, 
any T3E processor possesses 16 Megawords (128 Mbyte) of memory.
Our parallel code can manage roughly 100 K particles per processor,
roughly 1.8 times less than the serial code.
Therefore our code is not perfectly optimized as far as memory
is concerned. This is due partly to the fact that with respect to Dav\`e et al (1997)
we use twice the number of neighbours, and by the fact that we did not pose much restrictions
on th extension of the {\it ghost-tree} aiming at limiting the number of
communications.

\section{Adding new physics}
In this section we are going to present the collapse of a mix
of DM and baryons, with and without cooling. Firstly we show how cooling is implemented
in our Tree-SPH code, and how the integration of the energy equation is performed.
Then we show a simple model for the formation of a spiral galaxy, and then we discuss
the code performances when cooling is included.

\begin{figure*}
\centerline{\psfig{file=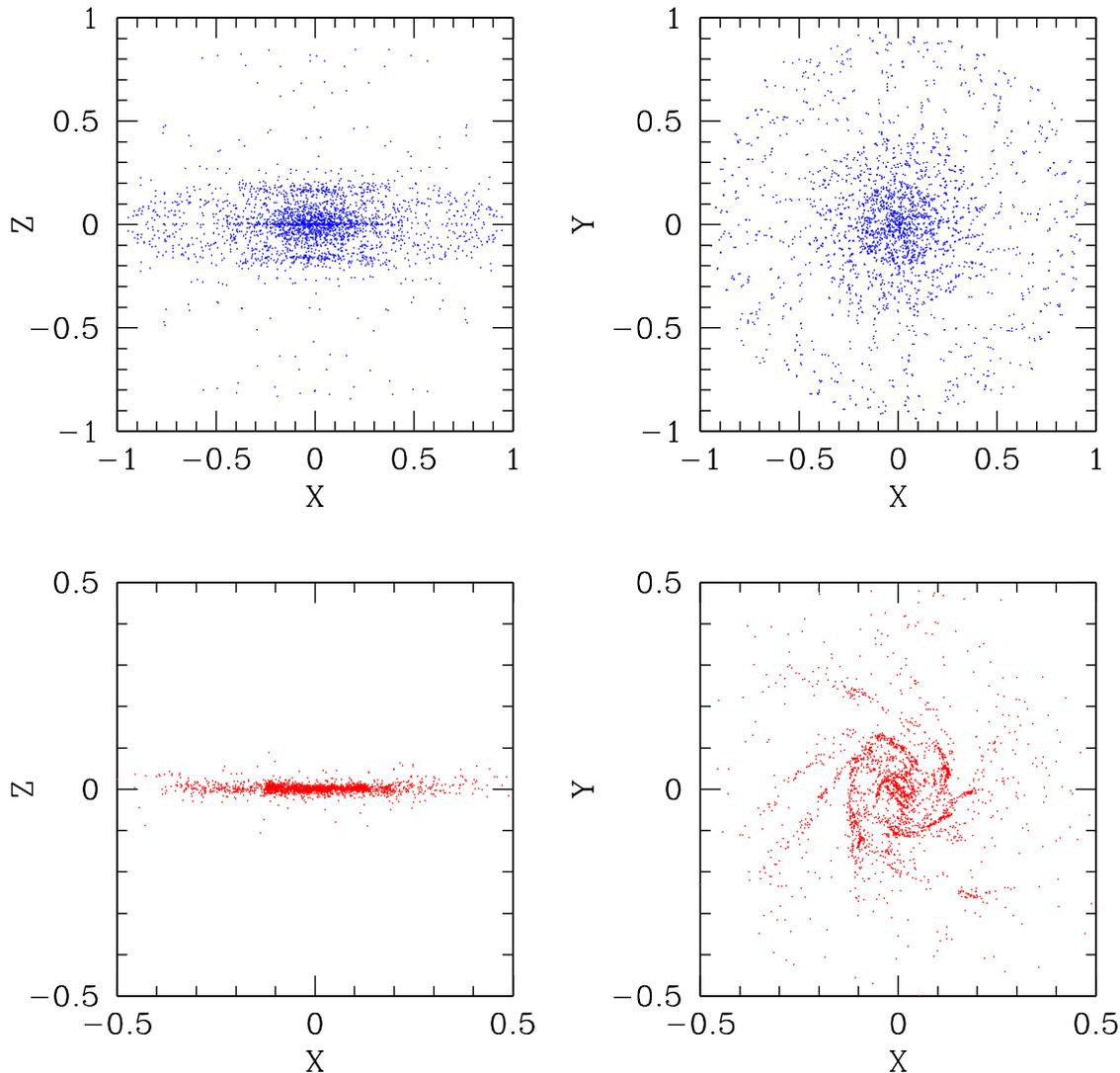,height=16cm,width=16cm}}
\caption{The collapse of a spherical mix of DM and gas
with cooling. Upper panels show
the final DM distribution, whereas lower panels show the gas
distribution. The $X-Z$ projection is on the right, while the $X-Y$ one
on the left.}
\end{figure*}

\subsection{On the integration of the energy equation}
The usual form of the energy equation  in SPH formalism is

\[
\frac{du_i}{dt}=\sum_{j=1}^N m_j \left
(\frac{\sqrt{P_{i}P_{j}}}{\rho_{i}\rho_{j}}
 + \frac{1}{2} \Pi_{ij}\right ){\bf v}_{ij} \cdot
\]

\begin{equation}
\frac{1}{2}
\left ( {\vec \nabla}_{i} W(r_{ij},h_i)+ \vec \nabla_i
W(r_{ij},h_j)\right ) + \frac {\Gamma - \Lambda_C}{\rho},
\label{eq_energy}
\end{equation}

\noindent
(Benz 1990; Hernquist \& Katz 1989). The first term
represents the heating or cooling  rate of mechanical nature, whereas
the second term  $\Gamma$ is  the total heating rate
from all sources apart from the mechanical ones, and the third term
 $\Lambda_C / \rho $ is the total
cooling rate by many physical agents (see Carraro et al. 1998a for
details).

In absence of explicit sources or sinks of energy the energy
equation
is adequately integrated using an explicit scheme and the Courant
condition for time-stepping (Hernquist \& Katz 1989).

The situation is much more complicated when considering cooling.
In fact, in real situations the cooling time-scale 
becomes much shorter 
than any other relevant  time-scale (Katz \& Gunn 1991),
and the time-step becomes considerably shorter than the Courant time-step. 
even  using the fastest  computers at disposal.
This fact makes it  impossible to integrate the complete system 
of equations (cf. Carraro et al. 1998a) adopting  as time-step the cooling
time-scale. 

To cope with this difficulty,
Katz \& Gunn (1991) damp the cooling rate to avoid too short
timesteps allowing gas particles to loose only half of their thermal energy
per timestep.
 
Hernquist \& Katz (1989) and Dav\'e et al. 1997 solve semi-implicitly eq.
(\ref{eq_energy}) using the trapezoidal rule,

\begin{equation}
u_{i}^{n+1} = u_{i}^{n} + \frac{1}{2} (dt_{i}^{n} \times e_{i}^{n} +
dt_{i}^{n+1} \times e_{i}^{n+1} ) .  
\label{trapez}
\end{equation}

The leap-frog scheme is used to update thermal energy, and the energy
equation,
which is nonlinear for $u_{j}$, is solved iteratively both at the
predictor and at the corrector phase.
The technique adopted is a
hybrid scheme which is a combination of the  bi-section and Newton-Raphson
methods (Press et al 1989). The only assumption is that at the predictor
stage, when the predicted $\tilde u_{i}^{n+1}$  is searched for, the terms
$u_{i}^{n+1}$ are equal to $u_{i}^{n-1}$.\\
 
Our scheme to update energy is conceptually the same, but differs in 
the predictor stage and in the iteration scheme adopted to solve
equation \ref{trapez}.

In brief, at the first time-step the quantity  $u^{n-1/2}$ is calculated
and for all subsequent time steps, the leap-frog technique, 
as in Steinmetz \& M\"uller (1993), is used: 
\vskip 0.2cm

(i)  We start with $u^{n}$ at $t^{n}$;\\

(ii) compute $\tilde u^{n}$ as 
 
\[
~~~~~~~~~~\tilde u^{n} = u^{n-1/2} + \frac{1}{2} t^{n} \times e^{n}
\]

\noindent
where $e_{i} = du_{i}/dt$.
This predicted energy, together with the predicted velocity is used to
evaluate the viscous and adiabatic contribution on  $e_{i}^{n+1}$.
In other words the predictor phase is calculated explicitly
because all the necessary quantities are available from the previous time
step $t^{n}$.\\

(iii) finally, derive $u^{n+1}$ solving the equation \ref{trapez}
iteratively (corrector phase) for both the predicted and old adiabatic and
viscous terms; \\

\begin{figure*}
\centerline{\psfig{file=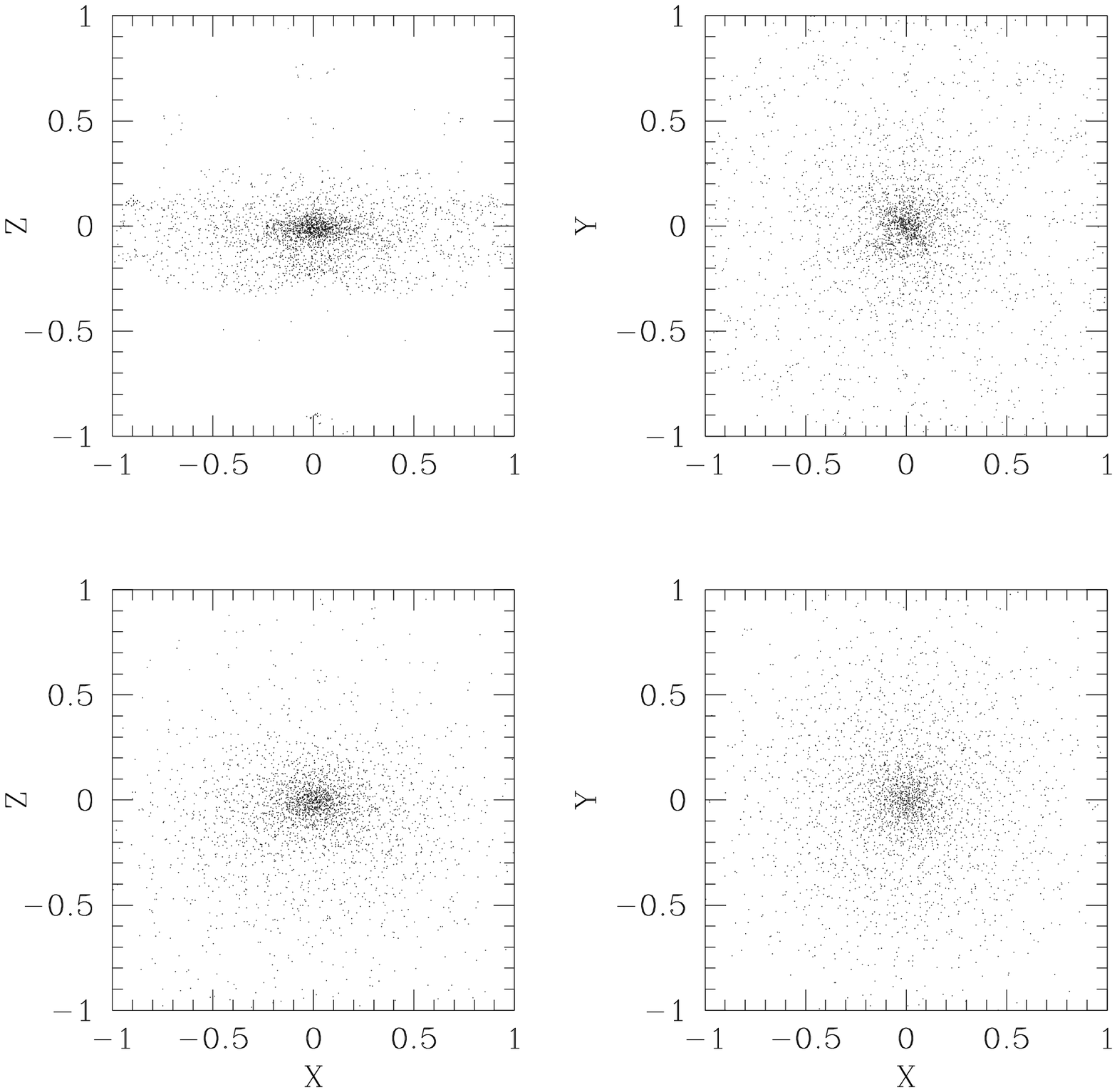,height=16cm,width=16cm}}
\caption{The collapse of a spherical mix of DM and gas without cooling. Upper
panels show
the final DM distribution, whereas lower panels show the gas
distribution. The $X-Z$ projection is on the left, while the $X-Y$ one
on the right.}
\end{figure*}

\noindent
In the corrector stage the integration of the equation \ref{trapez} 
is performed using the Brent method (Press et al 1989) instead of the
Newton-Raphson, the accuracy being fixed to a part in $10^{-5}$.
The Brent method has been adopted because it is better suited as
root--finder
for functions in tabular form (Press et al. 1989).\\
Radiative cooling is implemented as in Carraro et al 1998a.
We adopt cooling functions from Sutherland and Dopita (1993),
which are tabulated as a function of metallicity. Tables
are available for metallicity from $Z = 10^{-5}$ to $Z = 0.5$.\\

The parallel implementation of the radiative cooling does not present
any special difficulties. For each particle in fact there are already at disposal
all the necessary quantities (temperature and  density, basically)
to evaluate the amount of energy loss by cooling processes.
Only that in our code the overall efficiency is improved on
making the faster processors
to compute the amount of radiated energy not only for their own particles,
but also for the particles residing inside the slower processors.
This allows us to achieve a much better global load-balance (sse below).

\begin{figure*}
\centerline{\psfig{file=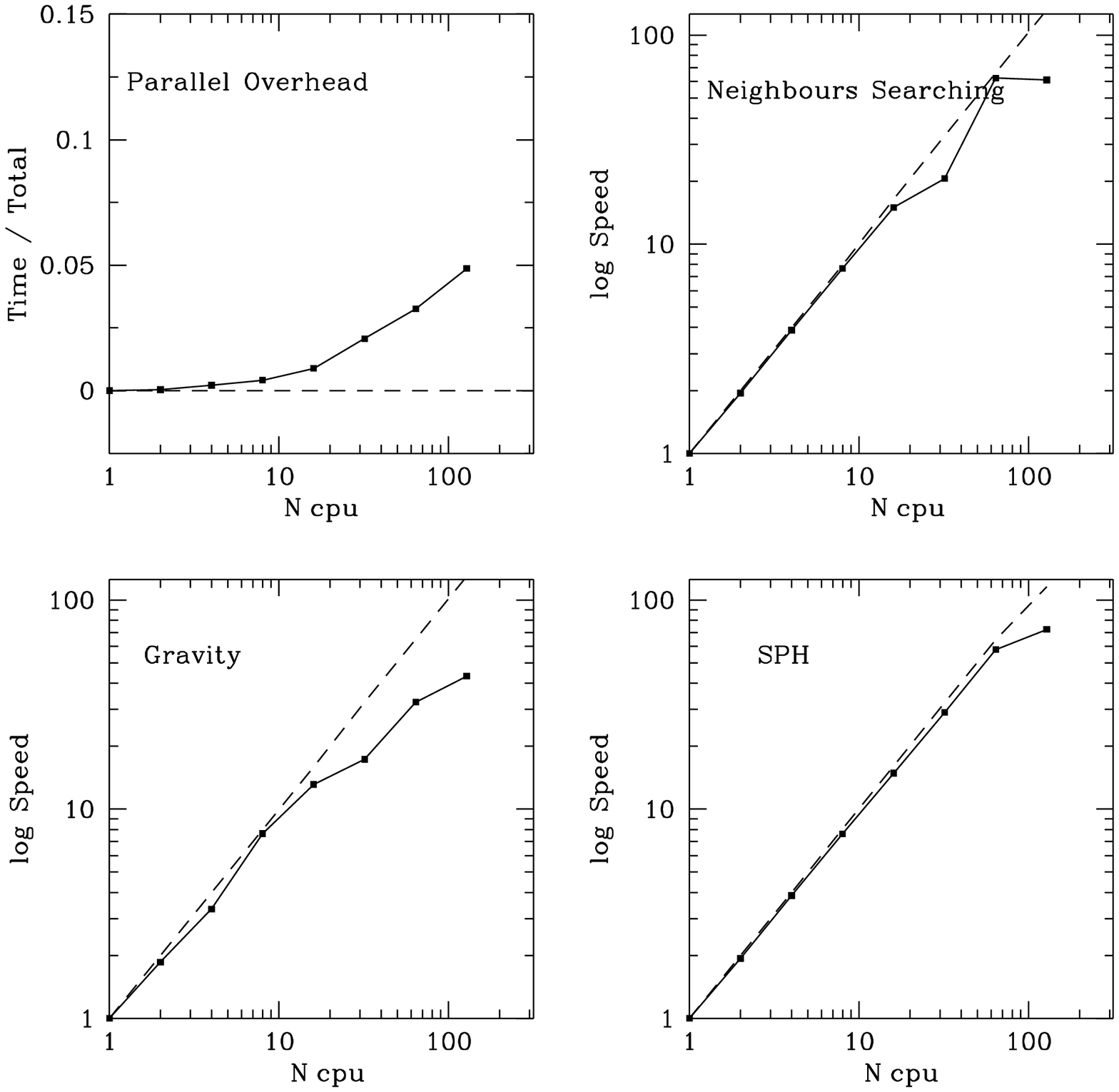,height=16cm,width=14cm}}
\caption{Scalability of different code sections (solid lines), as reported
in each panel, averaged
on 50 time-steps. Dashed lines indicate the ideal scalability.}
\end{figure*}

\begin{figure}
\centerline{\psfig{file=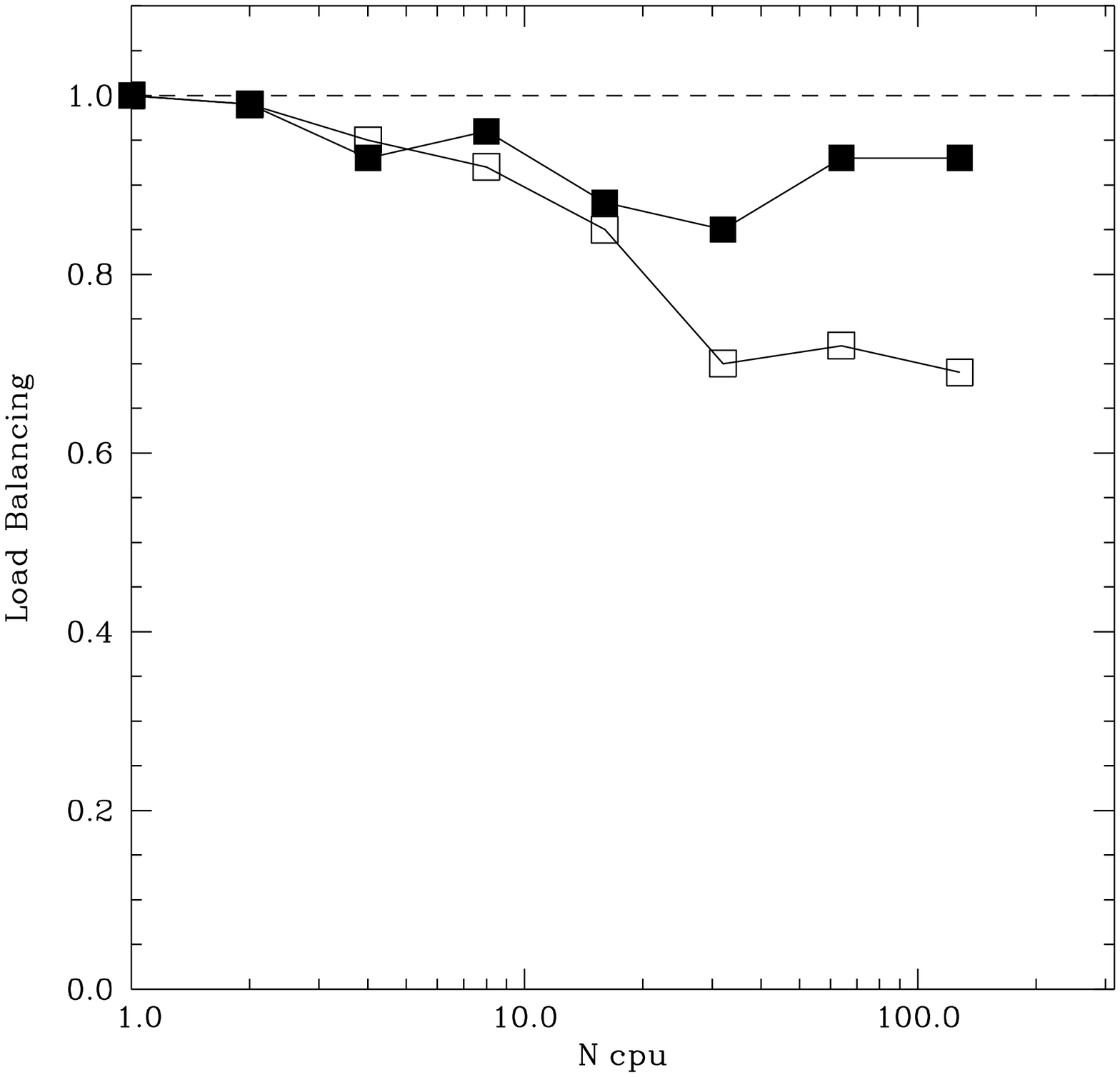,height=9cm,width=9cm}}
\caption{Overall code load-balance, averaged on 50 time-steps (solid
line) for the simulations described in Section~7.2
. Open squares refer to a simulation in which cooling is turned off,
whereas filled squares refer to a simulation in which cooling is switched on.
The dashed line indicates ideal scalability.}
\end{figure}

\begin{figure}
\centerline{\psfig{file=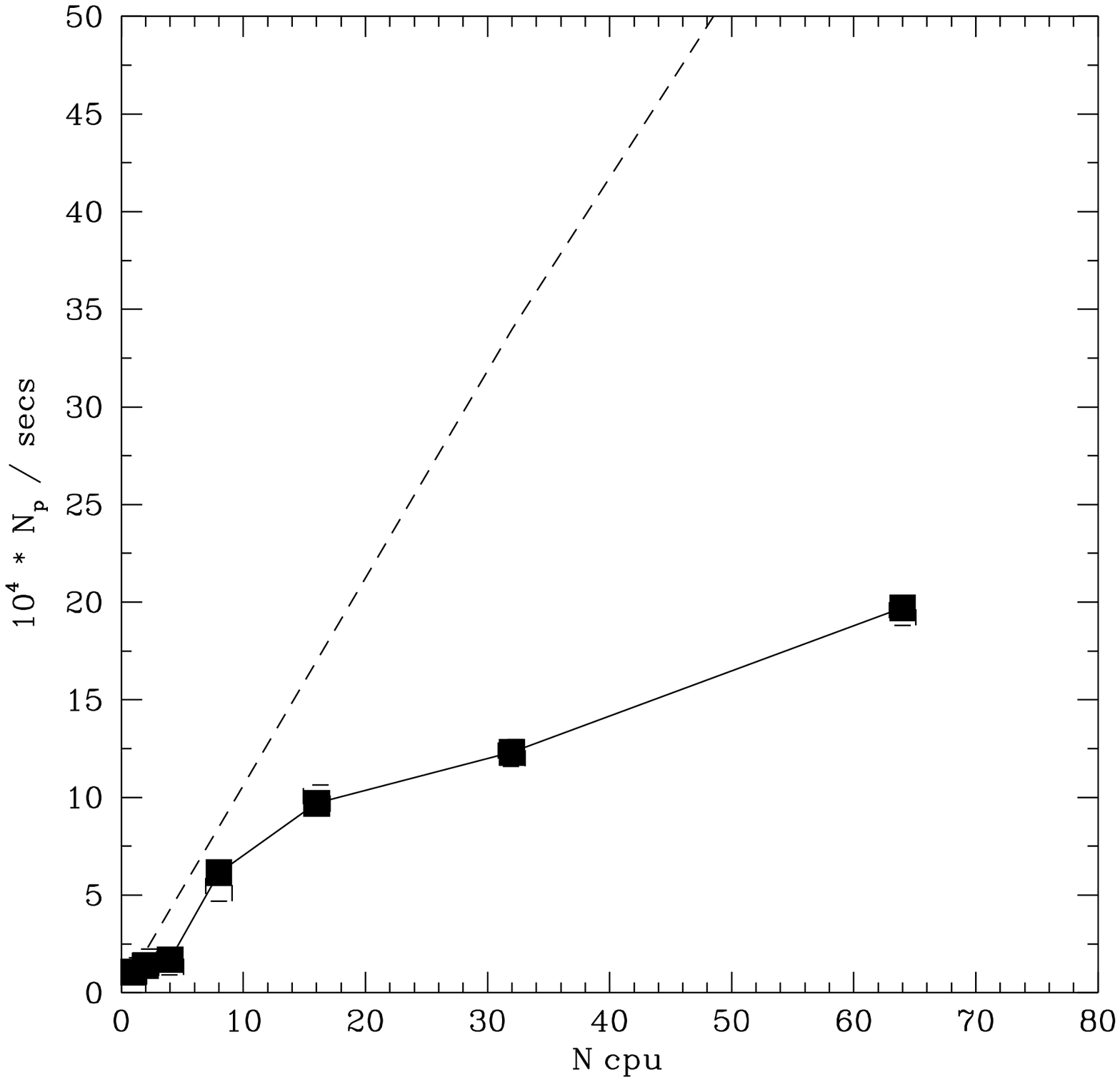,height=9cm,width=9cm}}
\caption{Overall code scalability, averaged on 50 time-steps (solid line)
for the simulations described in Section~7.2.
 Open squares refer to a simulation in which cooling is turned off,
whereas filled squares refer to a simulation in which cooling is switched on.
The dashed line indicates ideal load-balance.}
\end{figure}

\subsection{The Formation of a disk-like galaxy}
We consider a spherical DM halo whose density profile is

\begin{equation}
\rho(r) \propto \frac{1}{r} .
\end{equation}

\noindent
Although rather arbitrary,
this choice seems to be quite reasonable. Indeed DM halos emerging
from cosmological N--body simulations are not King or isothermal spheres,
but show, independently from cosmological models, initial fluctuations
spectra and total mass, a {\it universal} profile (Huss et al 1998).
This profile is not a power law, but has a slope $\alpha = dln\rho / d ln
r$ with $\alpha = -1$ close to the halo center, and $\alpha = -3$ at
larger radii. Thus in the inner part the adopted profile matches the
{\it universal} one. Moreover this profile describes a situation which is
reminiscent of a collapse within an expanding universe, being the local
free fall time a function of the radius (see for instance
the discussion in Curir et al 1993).\\   

DM particles are distributed in a regular cubic grid inside a sphere of
unitary radius (Carraro et al 1998a). The radial density profile in
eq.~(5) is realized stretching the initial grid (see eq.~20).\\
To mimic the cosmological angular momentum acquisition by tidal torque 
with the surrounding medium we
put the halo in solid body rotation around the
z-axis with an angular rotation velocity $\omega~=~0.5~sec^{-1}$,
which corresponds to $\lambda$
~=~ $0.09$, where

\[
\lambda = \frac{J |E|^{1/2}}{G M^{5/2}}
\]

\noindent
is the dimensionless spin parameter used to characterize the amount of
angular momentum in bound systems. $J$ is the total angular momentum
of the system, $E$ is the binding energy, $M$ is the total mass, and $G$
is the gravitational constant, kept equal to $1$ in our simulations.
$\lambda$ equal to $0.05$ is quite typical for halos emerging from
cosmological N--body simulations (Katz 1992, Steinmetz \& Bartelmann 1996).\\

For the simulations here described we used $10^{4}$ dark particles and
$10^{4}$ gas particles.
The softening parameter $\epsilon$ is computed as follows.
After plotting the inter-particles separation as a function of the distance
to the model center, we compute $\epsilon$ as the mean inter-particles
separation at the
center of the sphere, taking care to have at least one hundred particles
inside the softening radius (Romeo A. G. 1997). We
consider a Plummer softening parameter
and keep it constant along the simulation. It turns out ot be
$5 \times 10^{-3}$ .\\
To mimic the infall of gas inside the potential well of the halo,
we distribute gas particles  on the top of DM particles.
The baryonic fraction adopted is $f_{b}~=0.1~$, and gas particles
are Plummer--softened in the same way as the DM particles.\\
Under cooling and because of the velocity field of the halo,  
the gas is expected to settle
down in a rotating thin structure.\\
The results are shown in Fig.~8, where upper panels refer to the evolution of the dark 
component, whereas lower panels refer to the gas. Left panels show the evolution in the 
$X-Z$ plane, while right panels show the evolution in the $X-Y$ plane.
Due to the spin the dark halo flattens (upper-left panel), whereas the gas settles in a 
thin disk (bottom-left panel) which exhibits some spiral arms (lower-right panel).\\
These results are similar to those we have obtained with our scalar 
Tree-SPH code (Lia et al 1999).\\
This simulation is by no means aimed at producing a real disk, but only to show
that cooling is properly implemented.

A similar series of simulations, but without cooling, have been run, 
aiming at checking how the code performances change including new physics. 
In brief, we run the same simulation switching cooling off. The result is
shown in Fig~9. As expected without cooling, the gas component keeps round-shaped,
whereas DM gets flatter in the $X-Z$ plane (Carraro et al 1998a, Navarro \& White 1993).

\subsection{Benchmark}
As for the adiabatic collapse we run a series of  simulations of the formation of a disk
galaxy  at increasing number of processors.\\
We show the results in term of scalability and load balance in Figs~10, 11 and 12.
In Fig.~11 we present the scalability of 3 different code sections, namely 
the gravity computation, the neighbors searching and the SPH, measured as the wall-clock time spent
in a certain subroutine normalized to the global time-step, 
at increasing number of processors.
Looking at this figure, we conclude that the inclusion
of the cooling processes does not affect the scalability. The most significant deviation
is visible in the gravity part of the code, like in the case of the adiabatic collapse
(see Section~6), where the scaling starts deviating significantly
from the ideal one when using more than
32 processors.\\

On the other hand, the SPH and neighbors searching parts of the code
scale very well up to 64 processors.
The parallel over-head, which measures the time spent to synchronize the processors, increases
with the number of cpus, but always keeps below $10\%$.\\
In Figs~11 and 12 we compare the overall scalability and load-balance comparing a simulation
of a galaxy collapse with (see Fig.~8) and without cooling (see Fig.~9).
Filled squares represent the load-balance trend for the case of a collapse with cooling
turned on, whilst open squares refer to simulations in which cooling is turned off. \\

From Fig.~11,
it turns out that the inclusion of the cooling processes sensibly improves on the global
load-balance. 
This means that  our parallel scheme to calculate cooling partially alleviates 
computing time differences between fast and slow processors. 
This improvement is clearly due to the fact that the fastest processors 
compute the amount of energy radiated away by cooling also for the particles residing 
in the slowest processors. Going into some details, when a fast processor
has completed all the computation for its own particles, by using a SHMEM 
directive {\it shmem-int4-finc} it computes cooling for slower processors,
which are still working.\\
The two curves keep close
up to 16 processors, afterwards they start to deviate.\\
Fig.~12 presents the global code scalability for the two cases. No
difference can be outlined in the two series of simulations, showing
that the inclusion of cooling does not slow down the code, at odd with
what occurs in the serial code.
A clear departure from
the ideal scalability starts when using more than 16 processors. \\

We consider these results quite encouraging, 
also comparing our findings with Dav\'e et al (1997, see their Fig.~6) ones
as far as scalability is concerned.

\section{A spherically symmetric proto-cluster collapse}
Following Evrard (1988) step by step we simulate the turn-around and collapse
of a spherical perturbation in a flat cosmology ($\Omega = 1.0$) 
where the mass fraction 
of baryons over Dark Matter is $f_b = 0.1$.

\subsection{Initial conditions}
We generate an initial perturbation consisting of a single radial cosine wave
(Peebles 1982).\\
Given a uniform sphere of radius $x_t$, one perturbs comoving radii by

\begin{equation}
\vec x_0 \rightarrow \vec x_1 = x_0 \times (1 - \delta(x_0)/3)
\end{equation}

\noindent
and sets peculiar velocities according to growing linear theory

\begin{equation}
\vec v = -\frac{2}{3} H \vec x_0 \delta(x_0)
\end{equation}

\noindent 
where

\begin{equation}
\delta(x) = 0.5 \delta_0 \times [1 + cos(\pi x/x_t)] .
\end{equation}

\begin{figure*}
\centerline{\psfig{file=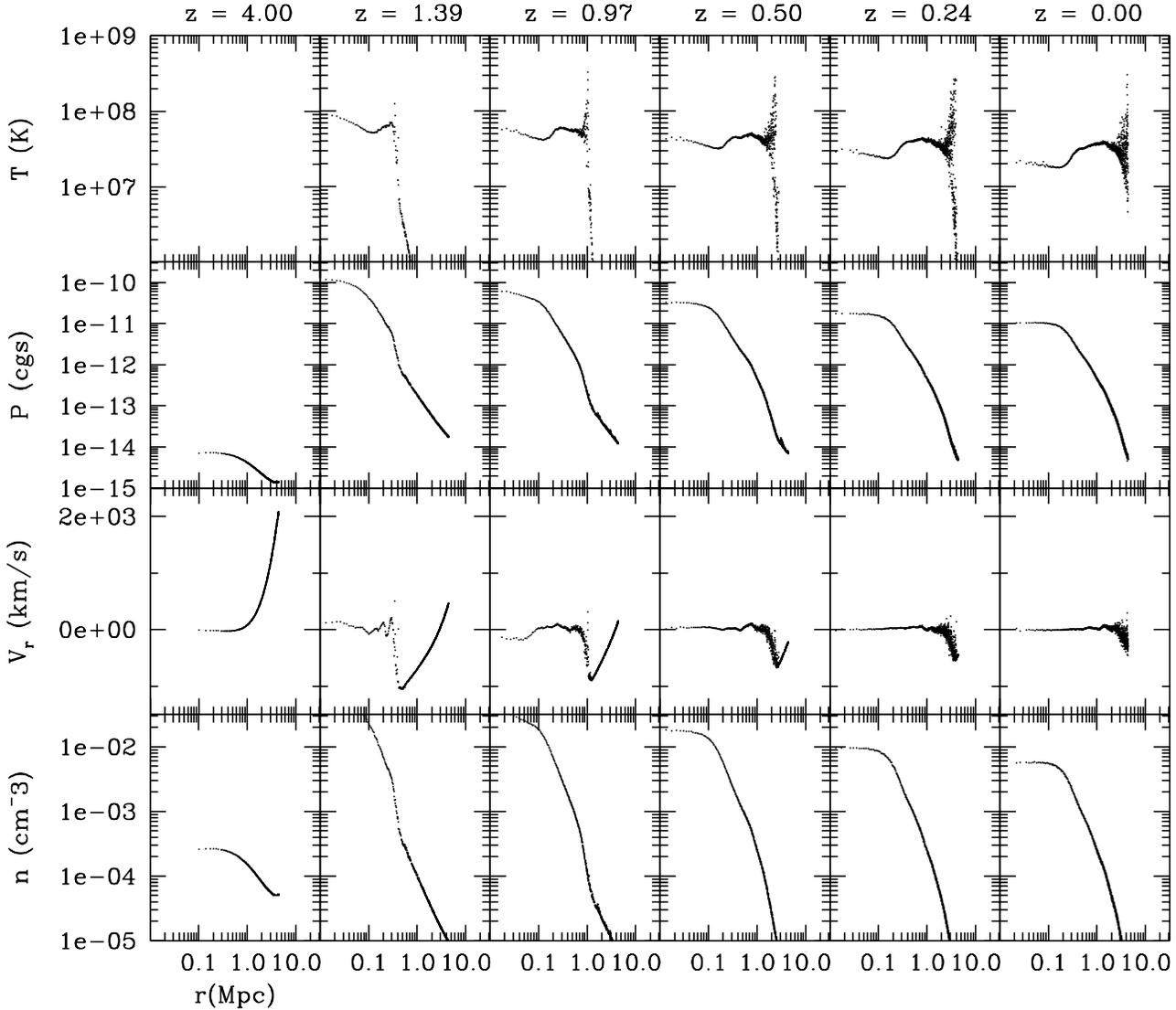,height=18cm,width=18cm}}
\caption{Evolution of the gas in the X-ray cluster collapse. Upper panels
show the evolution of temperature, middle panels pressure and radial velocity, lower
panels density. The results are shown for z = 4.00, 1.39, 0.97, 0.50,
0.24 and 0.00.}
\end{figure*}

\begin{figure*}
\centerline{\psfig{file=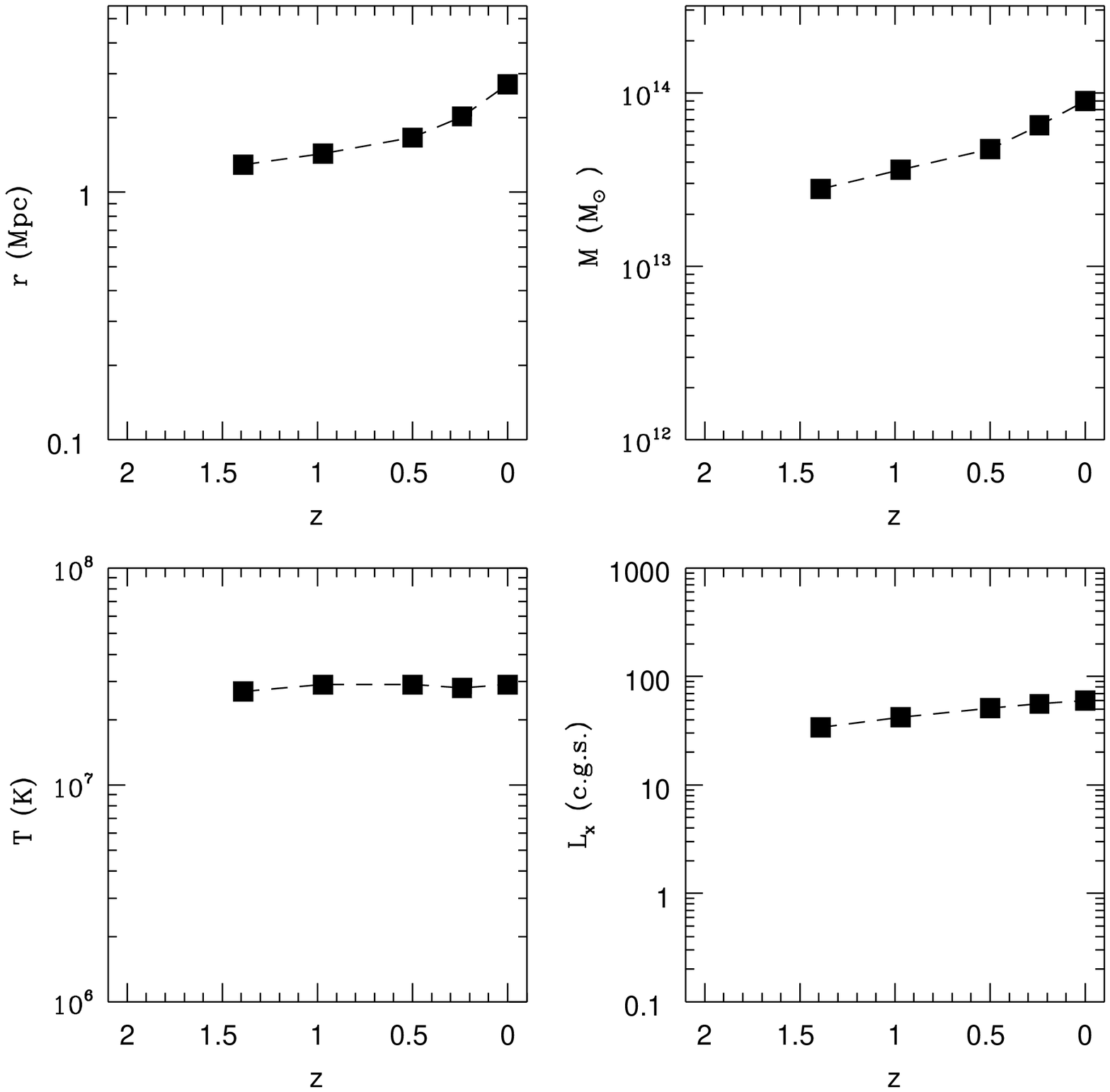,height=18cm,width=18cm}}
\caption{Redshift evolution of the core properties
of the simulated cluster. The core radius $r$, enclosed gas mass $M$, mean 
gas temperature $T$ and total X-ray luminosity ( in $10^{43} ~erg~sec^{-1}$ ) are shown. }
\end{figure*}

\noindent
The initial perturbation has $\delta_0 =1$ and it is scaled to a size appropriate
to rich cluster of galaxies. The cosmological parameter of the simulation are:

\begin{description}
\item [$\bullet$] mean density $\Omega$ = 1.0 ;
\item [$\bullet$] Hubble constant $H_0$ = 50$Km~sec^{-1}~Mpc^{-1}$ ;
\item [$\bullet$] baryon density (in units of the critical density) $\Omega_{b}$ = 0.1.
\end{description} 

\noindent
We used 100,000 gas particles and an equal number of collisionless DM particles within 
$x_t$. The model is evolved starting from $z = 4.0$, the total mass being
$M_t = 3 \times 10^{15} M_{\odot}$. The corresponding comoving radius is $x_t \approx
 22~ Mpc$ assuming $H_0 = 50~Km~sec^{-1}~Mpc^{-1}$.
The gas has an initial temperature of $10^{5}~^{o}K$.
Since the density perturbation is zero at $x = x_t$, the outer edges expand 
freely. 
We adopt a gravitational softening parameter $\epsilon \approx 5 ~kpc$.

\subsection{Results}
The simulation here presented was run using 32 processors and the full evolution required 
about 11 node-hours.\\
The detailed evolution of the baryonic component is shown in Fig~13, where we plot
in natural units (c.g.s.) the system number density, radial velocity, pressure and 
temperature at the same redshifts of Evrard (1988), namely $z$ = 4.00, 1.39, 0.97,
0.50, 0.24 and 0.00 .\\

The results agree well with Evrard (1988), who used about 3500 particles in total,
but are much closer to the 1D lagrangian 
finite difference results  reported by Evrard (1988) due to the much higher resolution
of our simulation. In particular the shock front is always very well defined,
and at $z = 0$ our profiles match fairly well the 1D ones (see Evrard, 1988 Fig~8),
especially at the center, which we succeed to resolve much better.\\

We finally compare the redshift evolution of the cluster core properties
to obtain a further check of the reliability of our results.
First of all we define the cluster 
core radius as the radius at which the mean interior density equals
a constant value
times the background density at that redshift.
The constant value has been chosen to be 170 to be consistent with  Evrard (1988).\\

For the output redshifts Fig.~14 shows the results for the redshift 
evolution of the core radius (in $Mpc$),
the enclosed baryonic mass (in $M_{\odot}$), the mean gas temperature (in $^{o}K$) and the total
X-ray luminosity, in units of $10^{43} erg ~sec^{-1}$.
We find a nice agreement with the 1D calculation presented by Evrard (see his Fig.~10).
In particular thanks to our higher resolution, we find good agreement also 
for the total X-ray luminosity.

\section{Summary and future perspectives}
In this paper we have presented a new parallel implementation of a Tree-SPH code,
realized by means of the SHMEM communications libraries for a 256 processors
T3E massively parallel computer.\\

We have shown that the code performs quite well against several well known tests,
like the adiabatic collapse of an initially isothermal gas sphere and the spherical
symmetric proto-cluster of galaxies collapse.\\

The qualitative results achieved and the 
global code load-balance and scalability are particularly encouraging.\\
The code is not portable at present, but it will be in the near future when
MPI 2 will be released. This way the code can be run also on different machines,
like for instance the IBM SP3, the new IBM parallel machine based on Power 3 cpus.
We believe that simulations of galaxy clusters formation which include 
cooling and star formation
are feasible using up to half a million particles of gas and an equivalent number
of collisionless particles. \\

We plan to use our code to address many different issues related to Galaxy
Formation, like for instance
the formation of elliptical galaxies, the interaction between Dark Matter and
baryons at galactic scale, and the structure and evolution of Damped Lyman $\alpha$
clouds.

\section*{Acknowledgments}
We thanks the anonymous referee for the careful reading of the manuscript
which led to an improvement of the presentation of this work.
The authors deeply acknowledge CINECA staff (in particular dr. Marco Voli)
for technical support and CNAA and SISSA for the computing time allocated
for this project. We acknowledge the enthusiastic support of 
our advisors, Proff. Luigi Danese and Cesare Chiosi, and many  
useful discussions with dr. Riccardo Valdarnini.
This work has been financed by several agency: the Italian Ministry of
University, Scientific Research and Technology (MURST), the Italian
Space Agency (ASI), and the European Community (TMR grant ERBFMRX-CT-96-0086).

\end{document}